\documentclass[letter,12pt]{article}

\usepackage{moreverb}

\usepackage{amsmath}
\usepackage{amsthm}
\usepackage{amssymb}
\usepackage{graphicx}
\usepackage{overpic}
\usepackage{subfigure}
\usepackage{color}
\usepackage{wrapfig}
\usepackage{comment}

\newtheorem{remark}{Remark}

\def\d{\partial}

\def\cE{\mathcal{E}}

\newcommand{\dS}{\displaystyle}

\newcommand{\bp}[1]{{\left(#1\right)}}

\newcommand{\sgn}{\mathrm{sign}\,}

\newcommand{\rM}[1]{{\mathrm{#1}}}

\newcommand{\pp}{\partial}

\def\sectionsplit{\newpage}

\def\dt{\triangle t}
\def\dx{\triangle x}

\def\sign{\text{sign}}

\def\sgn{\text{sgn}}

\sectionsplit{\newpage}

\begin{document}


\title{{\bf An explicit staggered-grid method for numerical simulation of large-scale natural gas pipeline networks}}

\author{V.~Gyrya\footnote{MS-B284, Los Alamos National Laboratory, Los Alamos, NM 87545, 
vitaliy$\_$gyrya@lanl.gov} 
and 
A.~Zlotnik\footnote{MS-B284, Los Alamos National Laboratory, Los Alamos, NM 87545, 
zlotnik@lanl.gov}
}

\maketitle

\begin{abstract}
	We present an explicit second order staggered finite difference (FD) discretization scheme
	for forward simulation of natural gas transport in pipeline networks.
	By construction, this discretization approach guarantees that the conservation of mass condition is satisfied exactly.
	The mathematical model is formulated in terms of density, pressure, and mass flux
	variables, and as a result permits the use of a general equation of state
	to define the relation between the gas density and pressure for a given temperature.
	In a single pipe, the model represents the dynamics of the density by propagation of a non-linear wave according to a variable wave speed.  We derive compatibility conditions for linking domain boundary values to enable efficient, explicit simulation of gas flows propagating through a network with pressure changes created by gas compressors.
	We compare Kiuchi's implicit method and an explicit operator splitting method with our staggered grid method, and perform numerical experiments to validate the convergence order of the new method.
	In addition, we perform several computations to investigate the influence of non-ideal equation of state models and temperature effects into pipeline simulations with boundary conditions over various time and space scales.
    %
	%

\end{abstract}

\maketitle


\section{Introduction}
\label{sec:intro}

Recent years have witnessed a renewal of interest in efficient methods for accurate simulation of transient flows through pipeline systems \cite{behbahani10,dorao11,woldeyohannes11,alamian12,grundel13a}.  The physical models and mathematical descriptions of fluid flow in pipelines under various regimes are well-understood \cite{wylie78}, and a variety of available numerical methods have been developed over decades \cite{thorley87,hudson06}.  In general, the Euler equations in one dimension are used to describe compressible fluid flow in a pipe using two hydrodynamic partial differential equations (PDEs) and an equation of state that relates gas density to pressure.  The Darcy-Wiesbach formula is used to approximate the nonlinear energy dissipation caused by three-dimensional turbulent flow within the pipe, while diffusive effects of turbulence are ignored.

A range of models and numerical methods is available for solution of the initial boundary value problem (IBVP) of gas flow through a pipeline system, depending on the time and space scales of interest.   A variety of high-fidelity, explicit methods are used to accurately approximate the effects of rapid variations in pressure and flow on time-scales corresponding to the speed of wave propagation \cite{zhou00,banda06,herty08}.  These methods solve the full isothermal Euler equations and account for nonlinear self-advection, so that the formation and propagation of shock waves can be captured.  They are typically second order accurate in regions where pressure and flux are continuous.  These methods are designed for accurate solution over short time scales and are impractical for simulation of large gas pipelines in the typical operating regime in which significant boundary changes occur on time scales of hours and longer.  Other studies have evaluated non-isothermal models \cite{osiadacz01,chaczykowski10,abbaspour08}.  On the other hand, several lower-fidelity, implicit methods are better suited to characterizing dynamics of large pipelines in the operating regime where pressure and mass flow change on time scales much longer than the rate of wave advection \cite{osiadacz84,kralik88,kiuchi94,grundel14}.  The terms in the Euler equations that may be omitted in this regime of interest have been characterized in previous empirical studies \cite{chua82,osiadacz84}, and the result is a nonlinear first order hyperbolic PDE system.   Such implicit methods are efficient, scalable, and adaptable to larger systems with complex network structure and nodal components such as gas compressors and regulators.  However, even slight changes on fast time-scales can present obstacles to stability of these methods \cite{kiuchi94}.

A major aim of recent work has been to bridge the gap between high-fidelity explicit methods that are computationally costly but stable for very fast transients \cite{abbaspour08}, and lower accuracy schemes that use simplified physics but can be applied to large-scale pipeline networks \cite{grundel14}.  One recently proposed method applies the well-known operator-splitting scheme to solve the simplified equations that describe one-dimensional dynamics of gas flow in networks of long pipes \cite{dyachenko16}.  This method accurately represents rapid, small amplitude changes  while resolving slowly-varying dynamics on the time scale of normal gas pipeline operations, and is extensible to complex pipeline networks with nodally-located time-varying compressors that boost gas flow into adjoining pipes.  Although it is unconditionally stable and second order accurate, the operator-splitting discretization is based on the propagation of characteristics through a uniformly-spaced grid, which requires the assumption of uniform and constant wave propagation speed, and thus an ideal gas model.  Because high pressure transmission pipeline flows exhibit highly non-ideal gas states, an explicit method for simulation of pipeline flows using non-ideal models is of significant interest.

In this paper, our aim is to develop a computationally efficient and explicit numerical method that is capable of simulating gas pipeline dynamics with non-ideal equation of state models.  We formulate a model for gas transport in a large-scale pipeline network.  For a single pipe, solution of an IBVP for this model could be viewed as representing the propagation of a non-linear wave in the density of the gas.  The non-linearity is present in the wave speed and in the non-linear dissipation term.  We present a new second-order accurate staggered grid discretization, with finite differences in space and time, for this model on a network and compare the resulting IVP solutions with other numerical schemes including that of Kiuchi \cite{kiuchi94}, explicit operator splitting \cite{dyachenko16}, and implicit lumped elements \cite{zlotnik15cdc}.

We focus in particular on the connection between the staggered grid approach and the implicit method of Kiuchi, which uses a standard (non-staggered) discretization where the staggered quantities are approximated using trapezoidal rule averages.  The implicit method of Kiuchi is unconditionally stable even when using an arbitrarily large time discretization step relative to the spatial discretization, although accuracy is reduced in this case.  Kiuchi's implicit method requires the solution of a system non-linear equations for each time step, which arise from the dissipation term and the generally complicated form of the compressibility factor in the non-ideal case.  Although our proposed explicit staggered finite difference discretization is restricted by a Courant-Levy stability condition, it employs simple time stepping that does not require solution of a large-scale implicit system.  Moreover, the data dependence on each time step is highly localized, which facilitates parallelization of computation of IBVP solutions for very large-scale (e.g. continental) pipeline networks.

In Section~\ref{sec:PDE formulation} we present a one-dimensional mathematical formulation for gas transport in a long pipe and various possibilities for modeling the dependence between gas density, pressure and temperature using a compressibility factor function.
We present the derivation of compressibility factor formulas in Appendix~\ref{sec:Z approximations}.
In Section~\ref{sec:pipe discretization}, we derive an explicit staggered grid discretization for non-ideal gas flow on a single pipe, then validate its second order accuracy, and derive a stability criterion for the grid steps.
In Section~\ref{sec:BC} we develop standardized compatibility rules for boundary conditions at junctions where two or more individual pipes are linked, so that the scheme can be applied to complex pipeline networks with gas compressors.
In Section~\ref{sec:numerics one pipe} we present numerical experiments on a single pipe to verify that the method is better than second order accurate, as well as to evaluate the significance of non-uniform temperature and non-ideal gas modeling for simulation of flows on various time and space scales.
In Section~\ref{sec:numerics on network} we examine a number of numerical experiments on a network of pipes using the proposed method as well as the operator splitting method and a lumped-element implicit method.
Finally, in Section~\ref{sec:conclusions} we conclude with a discussion and open questions for future work.


\section{Mathematical Modeling of Pipeline Dynamics}
\label{sec:PDE formulation}

The Euler equations in one dimension describe compressible fluid flow in a long pipe using two hydrodynamic PDEs and an equation of state that relates gas density to pressure \cite{hudson06}.  The Darcy-Wiesbach formula is used to approximate the nonlinear energy dissipation caused by three-dimensional turbulent flow within the pipe.  Because the effects of turbulent diffusion are ignored in that approximation, these equations can be considered as a phenomenological average of the three-dimensional Navier-Stokes hydrodynamic equations that captures effects over spatial scales much longer than the pipe diameter and over time scales longer than those required for dissipation of acoustic waves.  The rapid dissipation of turbulent effects in this regime of interest precludes the formation of shockwaves, which are strongly nonlinear acoustic events that develop and dispel over much shorter time scales as a result of significant abrupt changes \cite{wylie78,seleznev14}.

The Euler equations can be stated as two PDEs that describe the conservation of mass and momentum, respectively, as well as an equation of state.  The PDEs are
\begin{align}
	\d_t\rho+\d_x(v\rho) &= 0,
	\label{eq:euler0a}\\
	\d_t(\rho v) + \d_x(\rho v^2 + p)
	&=
	-\frac{\lambda}{2D}\rho v |v|-\rho g \sin \theta.
	\label{eq:euler0b}
\end{align}
The variables are gas velocity $v$, pressure $p$, and density $\rho$, defined on a domain $x\in [0,L]$ at time $t$.
The parameters are the friction factor $\lambda$, pipe diameter $D$, gravitational acceleration $g$, and pipe angle $\theta$. The terms on the right hand side of \eqref{eq:euler0b} aggregate friction and gravity effects.  We assume that gas pressure $p$ and density $\rho$ satisfy the equation of state relation
\begin{align}
	\label{eq:eos0}
	p&=Z(p,T)RT\rho,
\end{align}
where $R$ and $T$ are the ideal gas constant and temperature, respectively. Here $Z(p,T)$ is a compressibility factor that defines the deviation of the equation of state from the ideal gas law. We will assume that the equation \eqref{eq:eos0} can be uniquely solved for pressure as a function of density.

Equation \eqref{eq:euler0b} is valid in the regime when changes in gas consumption and injections are sufficiently slow to not excite propagation of sound waves.  Formally, the term $\pp_t(\rho v)$ on the left hand side is much smaller than $\pp_x(\rho v^2 + p)$. The ratio of the pressure gradient term $\pp_x p$ to the term $\pp_t(\rho v)$ is typically on the order of $1:0.01$, \cite{osiadacz84}. In addition, the flow velocities are much smaller than the nominal speed of sound,
$c=\sqrt{\overline{Z}RT}$, where $\overline{Z}$ is a nominal compressibility factor.  In this regime, the gas advection term $\pp_x(\rho v^2)$ can be omitted, even in comparison with the already small term $\dS\pp_t(\rho v)$.   For simplicity, we assume that the pipeline is level, thus ignoring the gravity term $\rho g \sin \theta$ on the right hand side of \eqref{eq:euler0b}.  We will usually assume that the gas temperature is uniform along the pipeline, so that $T$ is constant. One exception to this will be the Section~\ref{sec:temperature}, where we consider the effects of of heat generation by a compressor station.

With the above assumptions, \eqref{eq:euler0a} and \eqref{eq:euler0b} can be written in terms of (per area)
mass flux $\phi=\rho v$ (which has units kg/m$^2$/s) as
\begin{align}
 	\dS \pp_t\rho+\pp_x\phi & = 0,
 	\label{eq:euler1a} \\
  \dS \pp_t\phi + \pp_xp & = -\frac{\lambda}{2D}\frac{\phi |\phi|}{\rho}. 		
  \label{eq:euler1b}
\end{align}

The gas dynamics on a long pipe segment of a large gas transmission pipeline are commonly represented using \eqref{eq:euler1a}-\eqref{eq:euler1b}, subject to the initial and boundary conditions \cite{kiuchi94,zhou00}.  Simple boundary conditions on each end include one of $\rho(t,0)=\rho_{0}(t)$ or $\phi(t,0)=\phi_{0}(t)$, and one of $\rho(t,L)=\rho_{L}(t)$ or $\phi(t,L)=\phi_{L}(t)$.  More complicated boundary conditions can arise when we consider junctions of several pipes, as well as devices such as compressors and regulators that are used to actuate flow.

The compressibility factor $Z(p,T)$ is typically given as a formula with parameters that have been fitted to measured data obtained during early engineering studies \cite{benedict40,elsharkawy04}.  A simple and widely-used formula is the CNGA method \cite{menon05,lee66}, given by
\begin{align} \label{eq:cnga1}
	Z(p,T)& =\frac{1}{1 + \frac{a_1 p 10^{a_2G}}{T^{a_3}}},
\end{align}
where $a_1=344400$, $a_2=1.785$, and $a_3=3.825$ are nominal coefficients and $G$ is the gas gravity (with air $=1$ as reference).  Here $p$ is pressure in units of psig (gauge pressure), and $T$ is in units of $^\circ$R (degrees Rankine).
We consider a nominal gas gravity of $G=0.650784$ for 80\% methane and 20\% ethane.   The compressibility factor relation formula \eqref{eq:cnga1} can be used with the equation of state \eqref{eq:eos0} to create a bijection between density and pressure.  We let $P:\mathbb{R}_{\geq 0} \to \mathbb{R}_{\geq 0}$ denote a mapping from density to pressure acting $p=P(\rho)$, and let $P^{-1}$ be the inverse, acting $\rho=P^{-1}(p)$.  This mapping is derived for several cases of physical assumptions in Appendix~\ref{sec:Z approximations}.


\section{Numerical discretization scheme for a single pipe}
\label{sec:pipe discretization}

We now describe a numerical scheme for finite difference (FD) discretization of the dynamic equations (\ref{eq:euler1a}-\ref{eq:euler1b}) on a single pipe.  The proposed scheme is motivated by the fact that these PDEs reduce to the linear wave equation under the ideal gas assumptions for the constitutive law and zero friction, i.e.  $Z(p)\equiv 1$ and $f\equiv 0$. Therefore, we will adopt the classical second order accurate staggered discretization for a wave equation written in a mixed form.
This discretization could be considered a prototype of the classical Yee scheme for time domain computational electrodynamics \cite{Yee:1966}.

\begin{figure}[h!]
	\centering
	\includegraphics[width=.7\textwidth]{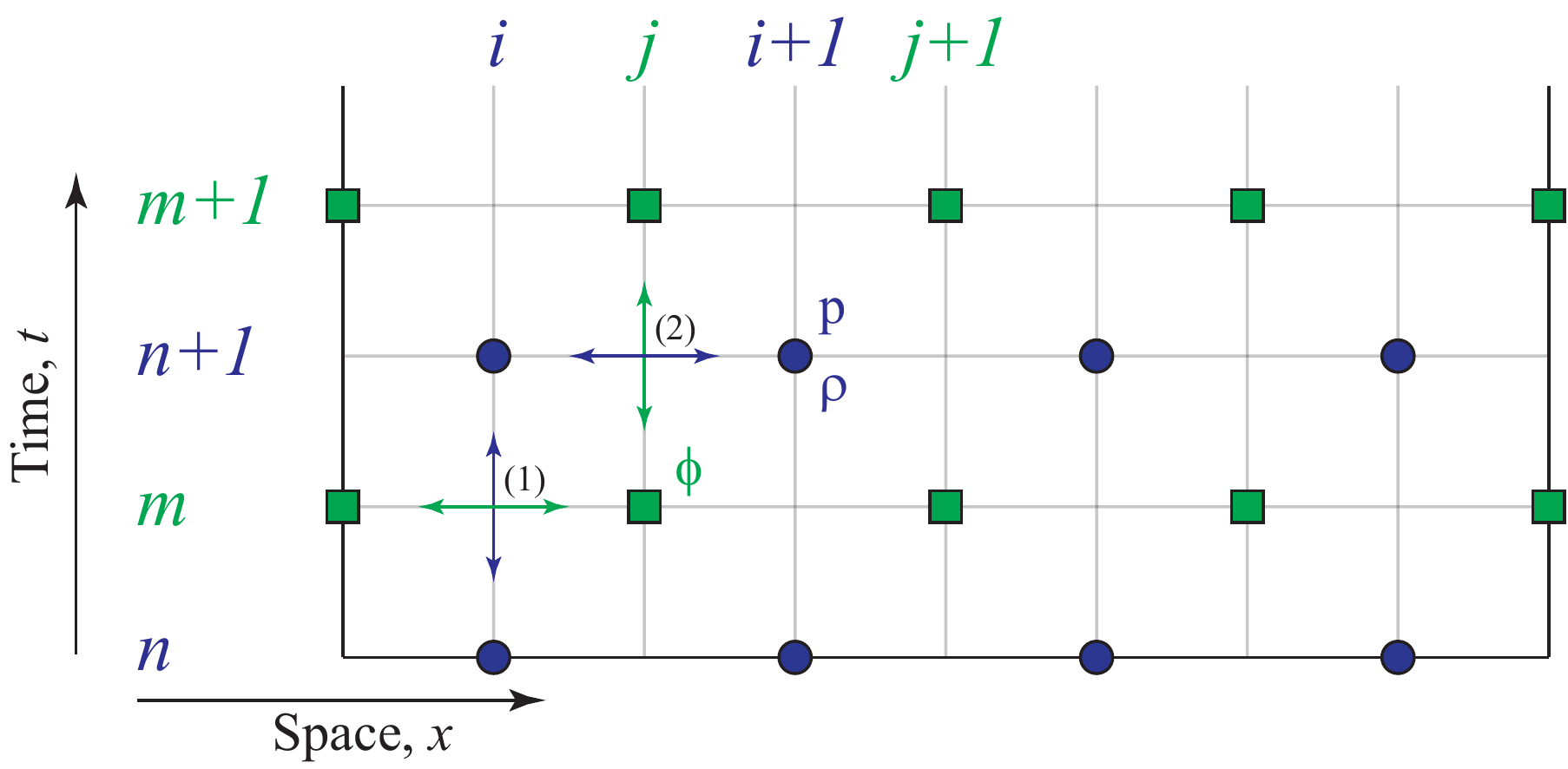}
	\caption{Illustration of the staggered grid discretization.
	The pressure, $p$, and density, $\rho$, are approximated at points $x_i$ and $t^n$
	(marked by circles).
	The flux, $\phi$, is approximated at points $x_j$ and $t^m$ (marked by squares).  }
	\label{fig:stag grid}
\end{figure}

Consider the staggered FD discretization shown in Figure~\ref{fig:stag grid}, where space grid points are indexed by two indices $i$ and $j=i+\tfrac{1}{2}$ and time layers are indexed by two additional indices $n$ and $m=n+\frac{1}{2}$. Variables $\rho$ and $p$ will be discretized  on the staggered grid in ``blue'' points (circles), while the flux variable $\phi$ will be discretized on the staggered grid in ``green'' points (squares).

\noindent
The equation \eqref{eq:euler1a} will be discretized,
at a point ($m,i$), as
\begin{equation}
	\label{eq:discr 1}
	\frac{\rho^{n+1}_{i}-\rho^{n}_{i}}{\dt}
	+
	\frac{\phi^{m}_{j} - \phi^{m}_{j-1}}{\dx}
	= 0.
\end{equation}
The equation \eqref{eq:euler1b} will be discretized
at a point ($n+1,j$), as
\begin{equation}
	\label{eq:discr 2}
	\frac{\phi^{m+1}_j-\phi^{m}_j}{\dt}
	+
	\frac{p^{n+1}_{i+1} - p^{n+1}_{i}}{\dx}
	= -\beta \frac{\left(\phi |\phi|\right)^{n+1}_j}{\rho^{n+1}_j},
\end{equation}
where $\beta=\lambda/(2D)$.  Observe that in the fraction on the RHS of \eqref{eq:discr 2}, $\rho$ and $\phi$ are evaluated between staggered grid points where their values are not available.  Therefore, we approximate the RHS of \eqref{eq:discr 2} by
\begin{equation}
\label{eq:approx of friction RHS}
	\frac{\left(\phi |\phi|\right)^{n+1}_j}{\rho^{n+1}_j}
	\approx
	\frac{\left(\phi |\phi|\right)^{m}_j+\left(\phi |\phi|\right)^{m+1}_j}
	{\rho^{n+1}_i+\rho^{n+1}_{i+1}}.
\end{equation}
By substituting \eqref{eq:approx of friction RHS}
in the equation \eqref{eq:discr 2} we get a
nonlinear implicit equation
\begin{equation}
	\label{eq:discr 2a}
	\phi^{m+1}_j
	+
	\beta\dt
	\frac{\left(\phi |\phi|\right)^{m+1}_j}
	{\rho^{n+1}_i+\rho^{n+1}_{i+1}}
	=
	\phi^{m}_j
	-
	\frac{\dt}{\dx} \left(p^{n+1}_{i+1} - p^{n+1}_{i}\right) 	
	-
	\beta\dt
	\frac{\left(\phi |\phi|\right)^{m}_j}
	{\rho^{n+1}_i+\rho^{n+1}_{i+1}}.
\end{equation}
Although the LHS is implicit in $\phi^{m+1}_j$, \eqref{eq:discr 2a} needs to be solved only pointwise, i.e. there is no coupling between $\phi^{m+1}_j$ and $\phi^{m+1}_{j+1}$. The function that needs to be inverted to solve for $\phi^{m+1}_j$ is
\begin{equation}\label{eq:invert function}
	F(x) = x
	+ 	
	\frac{\beta\dt}
	{\rho^{n+1}_i+\rho^{n+1}_{i+1}} x\, |x|
	=
	x \left(1	+ 	
	\frac{\beta\dt}
	{\rho^{n+1}_i+\rho^{n+1}_{i+1}} |x|\right).
\end{equation}
From the form of the RHS in \eqref{eq:invert function} one can see that the function $F(x)$ is a bijection on the real line. Indeed, both factors in the RHS of \eqref{eq:invert function} are increasing for $x>0$ and change sign at $x=0$. Hence, we make the following trivial, but useful observation
\begin{eqnarray}\label{eq:F x sign}
	\sign(F(x)) &=& \sign(x).
\end{eqnarray}
Using this observation we can write the inverse of $F(x)=y$ explicitly using the quadratic formula
\begin{equation}\label{eq:quadratic formula}
	\sign(y) ax^2+x-y=0 \qquad x
	=
	\sign(y)\frac{1}{2a}\left(-1\pm \sqrt{1+\sign(y)4ay}\right),
\end{equation}
where
\begin{equation}\label{eq:ay positive}
	a = \frac{\beta\dt}{\rho^{i+1}_n+\rho^{i+1}_{n+1}}>0
	\quad \text{and therefore} \quad
	\sign(y) a y = a |y|\geq 0.
\end{equation}
From \eqref{eq:F x sign} and \eqref{eq:ay positive} we conclude that only the positive solution
should be used in \eqref{eq:quadratic formula}.  Thus, the inverse of $F(x)$ has the closed form expression
\begin{equation}
\label{eq:inverse of F}
	F^{-1}(y)
	=
	\sign(y)\frac{-1 + \sqrt{1+4a|y|}}{2a}.
\end{equation}


\begin{remark}
	Note that in the limit of vanishingly small friction coefficient $\lambda$
	equation \eqref{eq:discr 2a} transforms into
	\begin{equation}
	\label{eq:discr 2 no A}
	\phi^{m+1}_j
	=
	\phi^{m}_j
	-
	\frac{\dt}{\dx} \left(p^{n+1}_{i+1} - p^{n+1}_{i}\right)
	=
	\phi^{m}_j
	-
	\frac{\dt c^2}{\dx} \left(\rho^{n+1}_{i+1} - \rho^{n+1}_{i}\right).	 	
	\end{equation}
\end{remark}

\subsection{Discretization order analysis}

%

\def\discrEqns{(\ref{eq:discr 1}-\ref{eq:discr 2})}

It is critical to observe that the discretization \discrEqns\ conserves mass in the system, as evident from  \eqref{eq:discr 1}.
Indeed, summing \eqref{eq:discr 1} for all space points $i$, one obtains
\[
	\sum_i \dx \rho^{n+1}_i
	=
	\sum_i \dx \rho^{n}_i
	+
	\dt( \phi^m_0 - \phi^m_N),	
\]
where $\phi^m_0$ and $\phi^m_N$ are the left-most and right-most values of the mass flux, respectively.

The discretization (\ref{eq:discr 1}-\ref{eq:discr 2}) is second order accurate in space and time, which can be validated by verifying that each of the terms in the discrete equations \discrEqns\ constitutes a second order accurate approximation to the corresponding terms in the PDEs (\ref{eq:euler1a}-\ref{eq:euler1b}).  
We provide this validation below by showing that the nonlinear term \eqref{eq:approx of friction RHS} is approximated with at least second order of accuracy.  Specifically, we aim to show that
\begin{equation}
\label{eq:friction order}
	\frac{\left(\phi |\phi|\right)^{n+1}_j}{\rho^{n+1}_j}
	=
	\frac{\left(\phi |\phi|\right)^{m}_j+\left(\phi |\phi|\right)^{m+1}_j}
	{\rho^{n+1}_i+\rho^{n+1}_{i+1}}
	+
	O(\dx^2,\dt^2).
\end{equation}
We refer to Figure~\ref{fig:stag grid} for the illustration of the staggered grid.

First, let us consider the numerator and the denominator of \eqref{eq:friction order}
independently, and show that they constitute a second order approximation.
Consider a Taylor expansion of $\rho(t,x)$ at points $(t^{n+1},x_i)$ and $(t^{n+1},x_{i+1})$, which yields
\begin{eqnarray}
	\label{eq:rho 1 avrg}
	\rho^{n+1}_i
	&=&
	\rho^{n+1}_j
	+
	(x_i\ \ \, -x_j)\rho'(t^{n+1},x_j)
	+
	\tfrac{1}{2}(x_i\ \ \, -x_j)^2 \rho''(t^{n+1},\xi_{i,j}),
	\\
	\label{eq:rho 2 avrg}
	\rho^{n+1}_{i+1}
	&=&
	\rho^{n+1}_j
	+
	(x_{i+1}-x_j)\rho'(t^{n+1},x_j)
	+
	\tfrac{1}{2}(x_{i+1}-x_j)^2 \rho''(t^{n+1},\xi_{{i+1},j}),
\end{eqnarray}
where where $\xi_{i,j}$ and $\xi_{i+1,j}$ are some points in the intervals $(x_i,x_j)$ and $(x_j,x_{i+1})$, respectively.  Noting that $(x_i -x_j) = \dx/2 = -(x_{i+1}-x_j)$ and taking an average of \eqref{eq:rho 1 avrg} and \eqref{eq:rho 2 avrg} yields
\begin{equation}
\label{eq:rho mean}
	\rho^{n+1}_j
	=
	\tfrac{1}{2}(\rho^{n+1}_i+\rho^{n+1}_{i+1})
	+
	O(\dx^2).
\end{equation}

For the numerator, we assume that $\phi$ does not change sign and treat $\phi\, |\phi|$ as one function.  As above, we then perform a Taylor expansion in time, similar to (\ref{eq:rho 1 avrg}-\ref{eq:rho 2 avrg}), and obtain
\begin{equation}
\label{eq:phi sqr mean}
	\left(\phi |\phi|\right)^{n+1}_j
	=
	\tfrac{1}{2}
	\left(\left(\phi |\phi|\right)^{m}_j+\left(\phi |\phi|\right)^{m+1}_j\right)
	+
	O(\dt^2).
\end{equation}
Substituting \eqref{eq:rho mean} and \eqref{eq:phi sqr mean} into the LHS of \eqref{eq:friction order} and performing a Taylor expansion in the correction terms $O(\dx^2)$ and $O(\dt^2)$ we obtain the RHS of \eqref{eq:friction order}.

\subsection{Stability analysis}

Although the discretization of the friction term \eqref{eq:approx of friction RHS} causes equation \eqref{eq:discr 2} to appear implicit, there is actually no spatial coupling between the $\phi^{m+1}$ terms as seen in \eqref{eq:inverse of F}, and therefore one should expect only conditional stability.  Furthermore, the nonlinearity of the PDE \eqref{eq:euler1b} precludes the performance of a Von-Neumann type analysis.  Instead, we will analyze what happens with a perturbation to a numerical solution on one time step and derive some error estimates that will include $\dt$, $\dx$ and solution properties.  The stability criterion that we derive below provides the condition under which we will be able to guarantee that the perturbation is not amplified so greatly as to dominate the solution.

To this end, we assume that $(\rho^n_i,p^n_i,\phi^m_j)$ and $(\rho^n_i+\tilde{\rho}^n_i,p^n_i+\tilde{p}^n_i,\phi^m_j+\tilde{\phi}^m_j)$ are two solutions of the numerical scheme,
and $(\tilde{\rho}^n_i,\tilde{p}^n_i,\tilde{\phi}^m_j)$ is the perturbation.  In order to maintain simplicity of notation, we introduce the discrete operators
\[
	\delta_x F^n_i:= \frac{F^n_{i+1/2} - F^n_{i-1/2}}{\dx}
	\qquad \text{and}\qquad
	\delta_t F^n_i:= \frac{F^{n+1/2}_{i} - F^{n-1/2}_{i}}{\dt}.
\]

Because the underlying PDE \eqref{eq:euler1a} and the corresponding discrete equation \eqref{eq:discr 1} are linear, the perturbation satisfies the similar discrete equation
\begin{equation}
\label{eq:eq1 pert}
	\delta_t\tilde\rho^{m}_i
	+
	\delta_x \tilde\phi^m_{i}
	=0.
\end{equation}
On the other hand, because the underlying PDE \eqref{eq:euler1b} and its discrete form \eqref{eq:discr 2}
are nonlinear, the perturbation (in general) is not itself a solution of the numerical scheme.  Instead,  we have
\begin{equation}
\label{eq:eq2 pert}
	\delta_t \tilde\phi^{n+1}_j
	+
	\delta_x \tilde{p}^{n+1}_j
	=
	-\beta
	\left(	
	\frac{(\phi+\tilde\phi)^2}{\rho+\tilde\rho}
	-
	\frac{\phi^2}{\rho}
	\right)^{n+1}_j,
\end{equation}
where we assumed that $\phi>0$ and $\phi+\tilde\phi$ in the domain of interest.  This is done mostly to simplify the notations, but also because this is where the physical assumptions of the model reductions are valid in order to obtain PDEs (\ref{eq:euler1a}-\ref{eq:euler1b}).  With regard to \eqref{eq:eq2 pert}, we note that $\tilde{p}$ does not satisfy the same equation as $p$.  Instead, we have
\begin{equation}
\label{eq:tp i n}
	\tilde{p}^{n}_{i}
	=
	P(\rho^{n}_{i}+\tilde\rho^{n}_{i})
	-
	P(\rho^{n}_{i})
	=
	P'(\rho^{n}_{i})\tilde\rho^{n}_{i}
	+
	O(\tilde\rho^{n}_{i})^2,
\end{equation}
where $P(\rho)$ is the point-wise dependence of $p$ on the value of $\rho$. Assuming the perturbations are small, we neglect higher order terms in \eqref{eq:tp i n}. The pressure term in \eqref{eq:eq2 pert} becomes
\begin{equation}
\begin{split}
	\delta_x \tilde{p}^{n+1}_j
	&=
	\left(\delta_x P'(\rho)^{n+1}_{j}\right)\ \tilde\rho^{n+1}_{j}
	+
	P'(\rho)^{n+1}_{j}
	\left(\delta_x \tilde\rho^{n+1}_{j}\right)
	.	
\end{split}
\end{equation}
By omitting higher order terms, the RHS of \eqref{eq:eq2 pert} simplifies to
\begin{equation}
\begin{split}
	\frac{(\phi+\tilde\phi)^2}{\rho+\tilde\rho}
	-
	\frac{\phi^2}{\rho}
	&=
	\frac{(\phi+\tilde\phi)^2}{\rho+\tilde\rho}
	\mp
	\frac{(\phi+\tilde\phi)^2}{\rho}	
	-
	\frac{\phi^2}{\rho}
	= \\ &=	
	(\phi+\tilde\phi)^2
	\left(\frac{1}{\rho+\tilde\rho} -\frac{1}{\rho}\right)
	+
	\frac{1}{\rho}\left((\phi+\tilde\phi)^2-\phi^2\right)
	= \\ &=
	(\phi+\tilde\phi)^2
	\left(-\frac{\tilde\rho}{\rho^2} +O(\tilde\rho^2)\right)
	+
	\frac{1}{\rho}\left(2\phi\tilde\phi + O(\tilde\phi^2)\right)	
	= \\ &=
	-\frac{\phi^2}{\rho^2}\tilde\rho
	+
	\frac{\phi}{\rho} \tilde\phi + O(\tilde\phi^2,\tilde\rho^2).		
\end{split}
\end{equation}
Again, neglecting higher order terms in $\tilde\phi$ and $\tilde\rho$
we obtain
\begin{equation}
\label{eq:eq2 pert 2}
	\left(
	\delta_t \tilde\phi
	+
	P'(\rho)
	\delta_x \tilde\rho
	\right)^{n+1}_j
	=
	\left(
	\left(
	\delta_x P'(\rho)
	+
	\beta \frac{\phi^2}{\rho^2} \right)\tilde\rho	
	-
	\beta
	\frac{\phi}{\rho}\tilde\phi	
	\right)^{n+1}_j.
\end{equation}

Summing up, the perturbations $(\tilde{\rho}^n_i,\tilde{p}^n_i,\tilde{\phi}^m_j)$ satisfy equations
\eqref{eq:eq1 pert} and \eqref{eq:eq2 pert 2}, which correspond to discretization of wave equation with a variable advection speed $\sqrt{P'(\rho)^{n+1}_{j}}$, and with the forcing term in the RHS of \eqref{eq:eq2 pert 2}.
In fact, $\sqrt{P'(\rho)^{n+1}_{j}}$ is the local speed of the non-linear equation as well.  Thus, the stability condition on the time step is formulated in terms of this wave speed:
\begin{align} \label{eq:stability_crit}
	\sqrt{P'(\rho)}\frac{\dt}{\dx}\leq 1.
\end{align}
Because the density changes as a function of space and time, we require this condition to be satisfied for the range of expected densities $\rho$, see e.g. method of frozen coefficients \cite{Strikwerda:2004}.


\section{Treatment of boundary conditions on a pipeline network}
\label{sec:BC}

In this section we describe a standardized approach for construction of pipeline network models with gas compressors for the purpose of simulating transient flows over large spatial scales.  This reduction extends the previous modeling work  \cite{zlotnik15cdc,zlotnik16ecc,misra15}.  We represent a pipeline network a metric graph with a system of PDEs on each edge, subject to time-dependent boundary conditions and nodal compressor actions, in a manner similar to what has been done for quantum graphs \cite{arioli17}.  The graph is defined by a set of nodes (or vertices) $\mathcal{V}$ and a set of directed edges $\mathcal{E}\subset V\times V$ with elements $(k,l)\in\mathcal{E}$ that represent pipes between the nodes $k,l\in\mathcal{V}$.  Every edge $(k,l)\in \mathcal{E}$ is associated with a diameter $D_{kl}$, length $L_{kl}$, cross-section area $S_{kl}$, and friction coefficient $\lambda_{kl}$, which constitute a metric on $\mathcal{E}$.  Gas flow through each pipe $(k,l)\in\mathcal{E}$ is characterized by the hydrodynamic equations \eqref{eq:euler1a}-\eqref{eq:cnga1} with solution denoted by $\rho_{kl}$, $p_{kl}$ and $\phi_{kl}$, and where $p_{kl}=P(\rho_{kl})$.  Here $P(\cdot)$ is the map defined in one of \eqref{eq:peq1} or \eqref{eq:peq2}.  For the pipe $(k,l)$, the sign of gas flow from node $k$ to node $l$ is denoted as positive by convention.  In our notation, we define the collection of pipes incoming to or outgoing from a node $l\in \mathcal{V}$ by
\begin{align}
\partial l&=\left\{ k\in \mathcal{V}\mid(k,l)\in \mathcal{E} \text{ or } (l,k)\in \mathcal{E}\right\} \subset \mathcal{V}. \label{eq:nodesout}
\end{align}
The balance of incoming and outgoing physical mass flows and injection of gas into the network must be preserved for each network node $l\in\mathcal{V}$. Such mass flow balance is maintained by applying the Kirchhoff-Neumann boundary conditions at network nodes, at which gas may also be injected or withdrawn from the system.  We denote the mass flow rate of gas withdrawn from the network at a node $l\in\mathcal{V}$ by the time-dependent function $q_l$.  The balance condition can then be written as
\begin{align} \label{eq:nodalflowbal0}
\sum_{k\in\partial_l} \sgn_{kl}S_{kl}\phi_{kl} = q_l ,  \quad \forall l\in\mathcal{V}.
\end{align}
where $\sgn_{kl}=1$ if flow $\phi_{kl}$ through $(k,l)\in\cE$ is denoted as incoming into $l$, and $\sgn_{kl}=-1$ if flow $\phi_{lk}$ through $(l,k)\in\cE$ is denoted outgoing from $1$.  Note that because mass flux $\phi_{kl}$ is expressed in per area units, while gas withdrawal $q_l$ is in mass flow units, the cross-section areas $S_{kl}$ are required to weight mass balance in the above equation.  That is, $\phi_{kl}(t,L_{kl})$ is the (per area) mass flux at the end of pipe $(k,l)$, and $S_{kl}\phi_{kl}(t,L_{kl})$ is the total mass flow into node $l$ from pipe $(k,l)$ if $\sgn_{kl}=1$.  Similarly, $\phi_{lk}(t,0)$ is the (per area) mass flux at the start of pipe $(l,k)$  and $S_{lk}\phi_{lk}(t,0)$ is the total mass flow out of node $l$ into pipe $(l,k)$ if $\sgn_{kl}=-1$.

\begin{figure}[h!]
  \[\includegraphics[width=.4\textwidth]{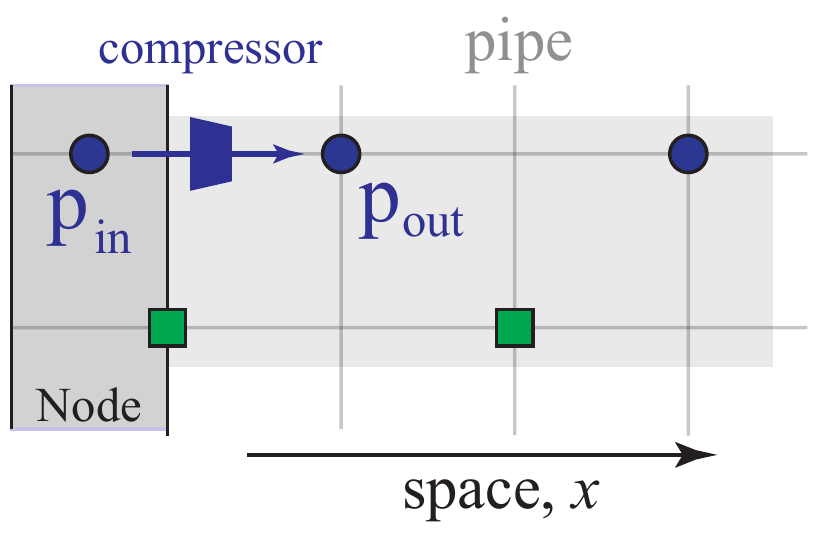}\]
\caption{Illustration of the model of a compressor.}
\end{figure}\label{fig:compressor}

Gas flow through the network is controlled by compressors that boost pressure while preserving mass flow.  In our model, such compressors are located at the interfaces between nodes and adjoining pipes, and are able to boost flow going from a node into an adjoining pipe, see Fig.~\ref{fig:compressor}.
This can be generalized by
\begin{equation}\label{eq:pressure gain BC}
	p_\text{out} = \alpha p_\text{in}
\end{equation}
where $\alpha\geq 1$ is a compression ratio that relates the suction (inlet) and discharge (outlet) pressures $p_\text{in}$ and $p_\text{out}$, respectively.  As with the other boundary conditions, we reformulate this condition in terms of density according to
\begin{equation}\label{eq:pressure gain for rho}
	P(\rho_\text{out}) = \alpha P(\rho_\text{in}).
\end{equation}
To model the action of such a compressor at the boundary of a pipe, we define auxiliary \emph{nodal} pressure variables $p_k$ and $p_l$ so that the function of compressors at the endpoints of a pipe $(k,l)$ can be described by
\begin{equation}\label{eq:generalized pressure gain}
p_{kl}(t,0)=\underline{\alpha}_{kl}(t)p_k(t) \qquad \text{and} \qquad p_{kl}(t,L_{kl})=\overline{\alpha}_{kl}(t)p_l(t),
\end{equation}
where $\underline{\alpha}_{kl}$ $\overline{\alpha}_{kl}$ are compression ratios that relate nodal pressures to boundary pressure at the inlet and outlet of the pipe, respectively.  In the method presented here, we assume that no gas is removed from the system by compression processes.  Complete boundary conditions can be provided for the pipeline network by specifying flow withdrawal or pressure at each node.  If flow $q_l$ removed from the network is specified, the pressure at that node is a dependent variable, and if pressure $p_l$ at a node is given, flow into or out from the network at that node is a free variable.  The pressure at at least one so-called ``slack'' node must be specified.   This endpoint boundary information, together with initial conditions, defines a well-posed IBVP for a pipeline network.

The equations \eqref{eq:nodalflowbal0} and \eqref{eq:generalized pressure gain} together constitute nodal compatibility laws that provide consistent boundary conditions for all the pipes in the network.  Next, we describe how the staggered grid discretization scheme can be applied to link pipes together with these compatibility conditions.

\subsection{Boundary conditions for a single pipe}
\label{sec:simple BC}

The most natural boundary conditions are on the values of the flux $\phi$, which are implemented for a single pipe simply by assigning the value to the flux at the ends.  Alternatively, the boundary conditions on pressure $p$ and density $\rho$ are reformulated at each time step by computing the appropriate value of the mass flux at that step that would result in those values of pressure and density, respectively.  Thus, for each pipe, each time step consists of the following sub-steps:
\begin{enumerate}
\item
	Compute the new value of the flux inside the domain.
\item
	Compute the appropriate values for the flux at the ends of the domain
	(left-most and right-most values).  This is the only step that incorporates the boundary conditions.
\item	
	Compute the values of the density (and hence pressure) at the next time step.
\end{enumerate}

In case that the boundary conditions specify values of mass flux at an endpoint of a pipe, step 2 above is straightforward.
However, if the boundary conditions specify the value of density at a pipe endpoint,
the appropriate value of the flux is found from the conservation of mass equation \eqref{eq:discr 1}.
For example, if $\phi^j_m=\phi_{kl}(t_m,L_{kl})$, i.e., $\phi^j_m$ gives the right boundary value for the mass flux on a pipe $(k,l)$, then we have at time $t_m$
\begin{equation}
	\phi^m_j
	=
	\phi^m_{j-1}
	-	
	\frac{\dx}{\dt}\left(\rho^{n+1}_i-\rho^{n}_i\right),
\end{equation}
where $\phi^m_{j-1}=\phi_{kl}(t_m,L_{kl}-\dx)$ is the internal value computed in step 1, $\rho^n_i=\rho_{kl}(t_n,L_{kl})$ is the density at a known time $t_n$,
and $\rho^{n+1}_i=\rho_{kl}(t_n,L_{kl})$ is specified by the boundary conditions.
In case the boundary conditions specify the value of pressure, this condition is reformulated in terms of the value of density using the map $P(\cdot)$,
and we proceed as above.

\subsection{Boundary conditions at a node connecting two pipes}
\label{sec:pressure gain BC}

In this section we describe incorporation of the multiplicative pressure gain condition \eqref{eq:pressure gain BC}, which is used to model the function of a gas compressor.

We first reformulate the condition on density as a condition on the mass flux.
Here we note that in order to satisfy mass conservation properties we must have a balance of mass flow through the joint.  This is expressed in terms of per area mass flux variables by
\[
	S_\text{in}\phi_\text{in} = S_\text{out}\phi_\text{out},
	\qquad \text{where} \qquad
	S_\text{in} = \pi D_\text{in}^2/4,
	\quad
	S_\text{out} = \pi D_\text{out}^2/4,
\]
where $S_\text{in},\ D_\text{in}$ and $S_\text{out}, D_\text{out}$
are the cross sectional areas and diameters of the pipes entering and leaving the joint (node), respectively.  However, we still can choose the value of the flux.  If $\phi^j_m$ is the value to be found, we then have the conditionss
\begin{eqnarray}
	\label{eq:pressure gain left}
	\phi^m_j
	&=&
	\phi^m_{j-1}
	-	
	\frac{\dx}{\dt}\left(\rho^{n+1}_i-\rho^{n}_i\right)
	\hspace{3mm}
	=
    \hspace{1mm}
	\phi^m_{j-1}
	-	
	\frac{\dx}{\dt}\left(P^{-1}(p_\text{in})-\rho^{n}_i\right)	
	,\\
	\label{eq:pressure gain right}
	\phi^j_{m}
	&=&
	\phi^m_{j+1}
	+	
	\frac{\dx}{\dt} \left(\rho^{n+1}_{i+1}-\rho^{n}_{i+1}\right)
	=
	\phi^m_{j+1}
	+	
	\frac{\dx}{\dt} \left(P^{-1}(p_\text{out})-\rho^{n}_{i+1}\right)
	,
\end{eqnarray}
where $\phi^m_{j-1}$ and $\phi^m_{j+1}$ are found during the step 1,
as described in Section~\ref{sec:simple BC},
$\rho^{n}_i$ and $\rho^{n}_{i+1}$ are known values,
and $\rho^{n+1}_i = \rho_\text{in}$ and $\rho^{n+1}_{i+1} = \rho_\text{out}$
are the values related by \eqref{eq:pressure gain for rho}.

\begin{figure}[h!]
\centering
\includegraphics[width=.8\textwidth]{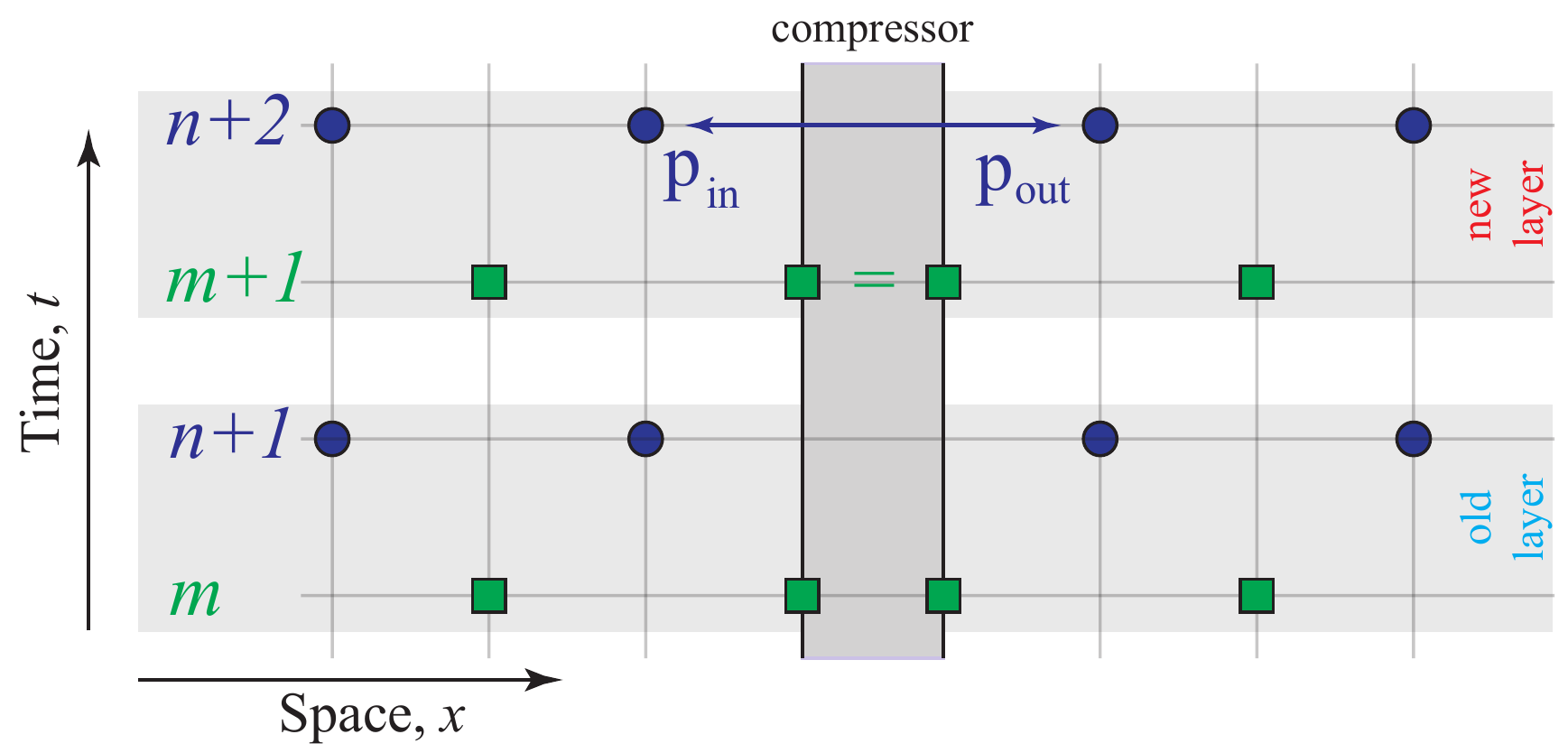}
\caption{Illustration of pressure gain boundary conditions
for two pipes adjacent to the node, representing two pipes joined by a compressor.}
\end{figure}

Thus, the pressure gain boundary condition \eqref{eq:pressure gain BC}
is implemented by solving
(\ref{eq:pressure gain for rho}-\ref{eq:pressure gain right})
for the value of the flux $\phi^m_{j}$,
for which we have the two expressions
\begin{eqnarray}
	\label{eq:pressure gain left 2}
	\phi^m_j
	&=&
	\phi^m_{j-1}
	-	
	\frac{\dx}{\dt}\left(P^{-1}(p_\text{in})-\rho^{n}_i\right)	
	,\\
	\label{eq:pressure gain right 2}
	\phi^m_{j}
	&=&
	\phi^m_{j+1}
	+	
	\frac{\dx}{\dt} \left(P^{-1}(\alpha p_\text{in})-\rho^{n}_{i+1}\right)
	.
\end{eqnarray}
From (\ref{eq:pressure gain left 2}-\ref{eq:pressure gain right 2})
we can obtain a solution for $p_\text{in}$ using the relation
\begin{equation}\label{eq:pressure gain pin}
	\left(
	P^{-1}(\alpha p_\text{in})+P^{-1}(p_\text{in})
	\right)
	=
	\left(\rho^{i}_n+\rho^{i}_{n+1}\right)	
	+
  \frac{\dt}{\dx}	
	\left(\phi^j_{m-1} - \phi^j_{m+1}\right)	
	.	
\end{equation}
Note that the solution of \eqref{eq:pressure gain pin} is unique
when the function $P(\cdot)$ is one-to-one.
In particular, this is the case when $P(\cdot)$
(and hence $P^{-1}(\cdot)$) is a monotone function.
For example, in the case \eqref{eq:eos2}
the equation \eqref{eq:pressure gain pin}
is a quadratic equation for $p_\text{in}$ that has a unique positive solution.

\subsection{Boundary conditions for a joint with multiple pipes}
\label{sec:presure gain multiple}

In this section, we consider a generalization of the pressure gain boundary condition discussed in the previous section to the case of three and more pipes.  For generality, we also assume that compressors are present at the endpoints of every pipe $(k,l)\in\mathcal{E}$.  Suppose that multiple pipes are adjoined at a node $k\in\mathcal{V}$, which is associated with a unique pressure $p_k$.  Thus, the pressure gain condition will be written in the form
\begin{equation}\label{eq:generalized pressure gain 1}
	p_{kl}(t,0)=\underline{\alpha}_{kl}(t)p_k(t) 
	\qquad \text{and} \qquad 
	p_{kl}(t,L_{kl})=\overline{\alpha}_{kl}(t)p_l(t),
\end{equation}
where the boost ratios $\underline{\alpha}_{kl}$ and $\overline{\alpha}_{kl}$ represent the actions of compressors propelling gas into the pipe from the endpoints.  Here, $p_{kl}(t,0)$ and $p_{kl}(t,L_{kl})$ represent the pressures at the endpoints of the pipe $(k,l)$, and $p_k$ and $p_l$ represent the pressures at the nodes on the other side of the compressors located at the endpoints.  In this general setting, at least one of $\underline{\alpha}_{kl}$ and $\overline{\alpha}_{kl}$ are set to unity at any given time, because both compressors would not be used simultaneously.

The conservation of mass condition for the joint is given by the flow balance equation \eqref{eq:nodalflowbal0}.
As done in equation \eqref{eq:pressure gain left 2}, applying the discretization to the conservation of mass equation,
we obtain
\begin{equation}\label{eq:pressure gain BC generalized}
	\sgn_{kl}
	\left(\phi^m_{kl} - \phi^m_{kl-}\right)
	=	
	\frac{\dx_{kl}}{\dt}\left(\rho(\alpha_{kl}\, p_k^{n+1})-\rho^{n}_{kl}\right),
\end{equation}
where $\phi^m_{kl-}$ is the flux at the node adjacent to the end node $\phi^m_{lk}$ (equal to $\phi_{kl}(t_m,L_{kl}-\dx_{kl})$ if $\sgn_{kl}=1$ and $\phi_{kl}(t_m,\dx_{kl})$ if $\sgn_{kl}=-1$);
$\rho^{n}_{kl}$ is the known value of density at the $k$ node end of the pipe $(k,l)$
and $\dx_{kl}$ is the spatial discretization step for the pipe $(k,l)$.
Multiplying \eqref{eq:pressure gain BC generalized} by $S_{kl}$ and
summing up over all pipes going into the node
and using the balance of flows equation \eqref{eq:nodalflowbal0},
we get the new version of \eqref{eq:pressure gain pin}:
\begin{equation}\label{eq:pressure gain pin generalized}
	\sum_{k\in\partial l}
	\frac{\dx_{kl}}{\dt}S_{kl}P^{-1}(\alpha_{kl}\, p_l^{n+1})
	=
	\sum_{k\in\partial l}
	\frac{\dx_{kl}}{\dt} S_{kl}\rho^{n}_{kl}
	+
	q_l
	-
	\sum_{k\in\partial l}
	\sgn_{kl}
	S_{kl}
	\phi^m_{kl-}.
\end{equation}
This is an equation on the node pressure $p_l^{n+1}$.
All other quantities in \eqref{eq:pressure gain pin generalized}
are known.
Once $p_l^{n+1}$ is solved for,
from \eqref{eq:pressure gain BC generalized} one finds the values of $\phi^m_{kl}$, which are used as the boundary conditions on each of the pipes separately.

\begin{remark}
In this paper we consider only the case
where the time discretization step $\dt$ is the same for all pipes $(k,l)$,
while the spatial discretization step $\dx_{kl}$ may vary.
Also note that on each pipe we have to satisfy an appropriate CFL condition
relating $\dx$, $\dt$ and the wave speed.
\end{remark}


\section{Numerical experiments on a single pipe}
\label{sec:numerics one pipe}

In this section we perform a number of numerical experiments
designed to compare and contrast various models.
One of the comparisons we perform is based on the choice of the constitutive law
for $Z(p)$.
The two models to be considered are
\begin{eqnarray}
	\label{eq:Z ideal gas}
	Z(p)&\equiv& 1\, \hspace{14mm}\text{ -- ideal gas,}\\
	\label{eq:Z non ideal gas}
	Z(p)&\equiv& \frac{1}{b_1 + b_2 p} \hspace{3mm}\text{ -- non-ideal gas,}
\end{eqnarray}
where $b_1=1.00300865$, and $b_2=2.96848838\times 10^{-8}$.

\subsection{Verification of the order of the method}
\label{sec:order verification pipe}

In order to verify the order of convergence of the method
we considered a single initial value problem on one pipe
(with length $L=10^4$ meters)
solved numerically with different resolutions
($\dt=1,3^{-1},3^{-2},\dots,3^{-5}$)
but fixed time and space resolution
$\dx/\dt=454.55$ ratio.
The reason for this choice of the resolution ratio
is to keep it larger than the maximum possible speed of sound in the pipe.
As a reference we also use a numerical solution
with a much higher resolution ($\dt=3^{-6}$).

We chose to do a $3\times$ refinement instead of the more standard $2\times$
so that the discretization points for the pressure $p$ and the density $\rho$
for a coarser descritization are a subset of fine discretization.
For $2\times$ refinement, this is not the case.

We also note that in order to set up the discrete initial value problem
we need to specify the values of $\rho$ and $\phi$
which are defined at different times ($t=0$ and $t=\dt/2$).
We set up all problems with the same initial values of density
\[
	\rho(t=0,x)
	=
	\overline{\rho}
	\left(1	- \frac{0.2}{\pi}
	\arctan\left(\frac{10(x-\tfrac{L}{2})}{L}\right)\right),
	\qquad
	\overline{\rho} = 56.817,
\]
which has an $\arctan()$ step profile centered at the center of the pipe
and shifted upward by a mean value $\overline{\rho}$.
We then set up the fine-grid reference problem with
the initial value for the flux
\[
	\phi(t=\dt_\text{fine},x)
	=
	c_\text{ref}
	\overline{\rho}
	\left(1	- \frac{0.2}{\pi}
	\arctan\left(\frac{10(x-\tfrac{L}{2}-c_\text{ref}\dt_\text{fine})}{L}\right)\right),
	\quad
	\left\{
	\begin{array}{l}
	\dt_\text{fine}=3^{-6},\\
	c_\text{ref} = 377.9683.	
	\end{array}
	\right.
\]
The initial conditions for the flux $\phi$ on coarse grids are
obtained from the solution for $\phi$ on the fine grid
(by restriction)
at the time $\dt_\text{coarse}/2$.

These initial conditions were selected because in the case of a simple
ideal gas model $Z(p)\equiv 1$ and zero friction $f=0$
we get a traveling wave solution moving in positive direction with
the speed $c_\text{ref}$.

\begin{figure}[h!]
	\[	
	\includegraphics[width=.49\textwidth]{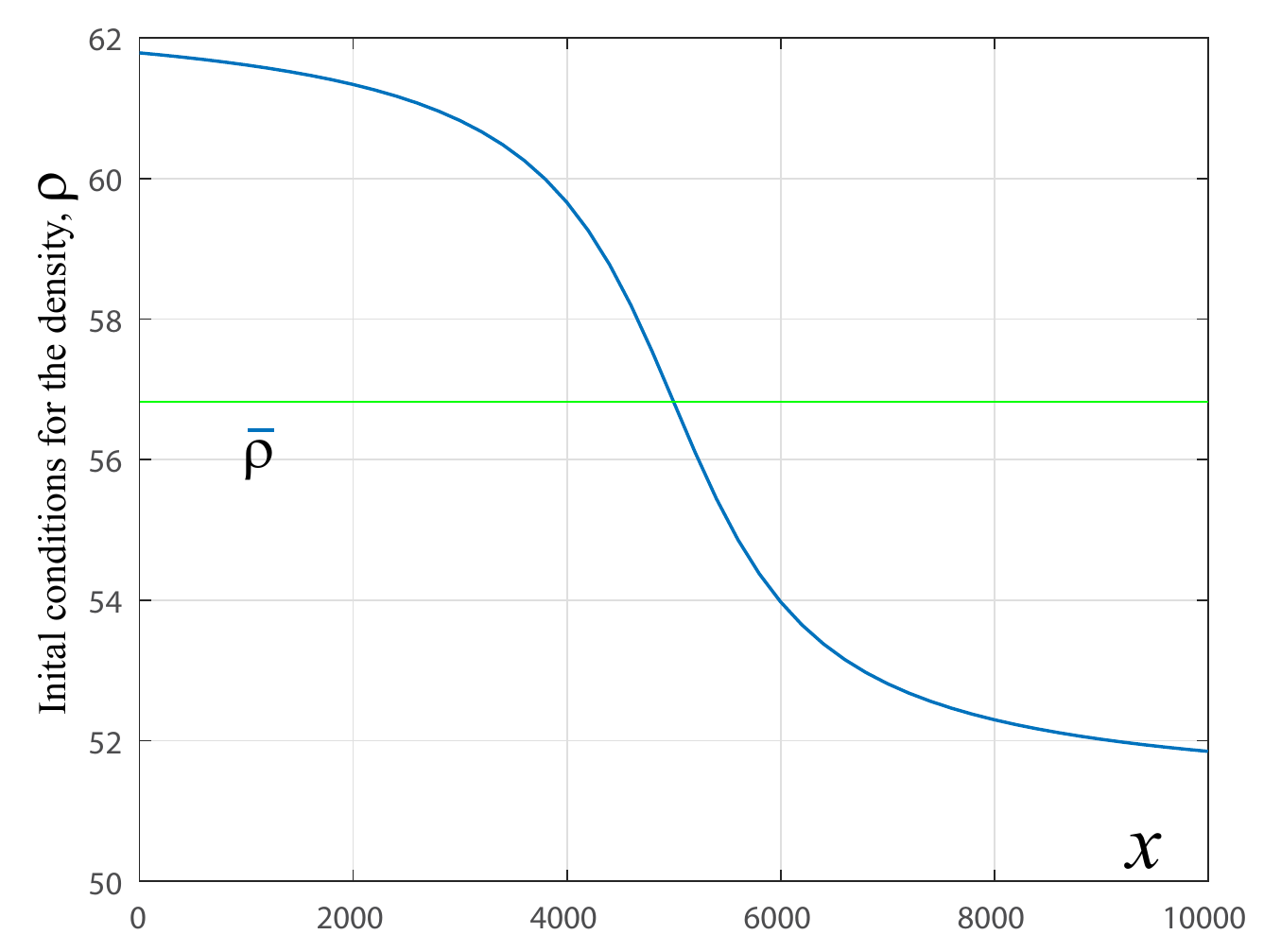}
	\includegraphics[width=.49\textwidth]{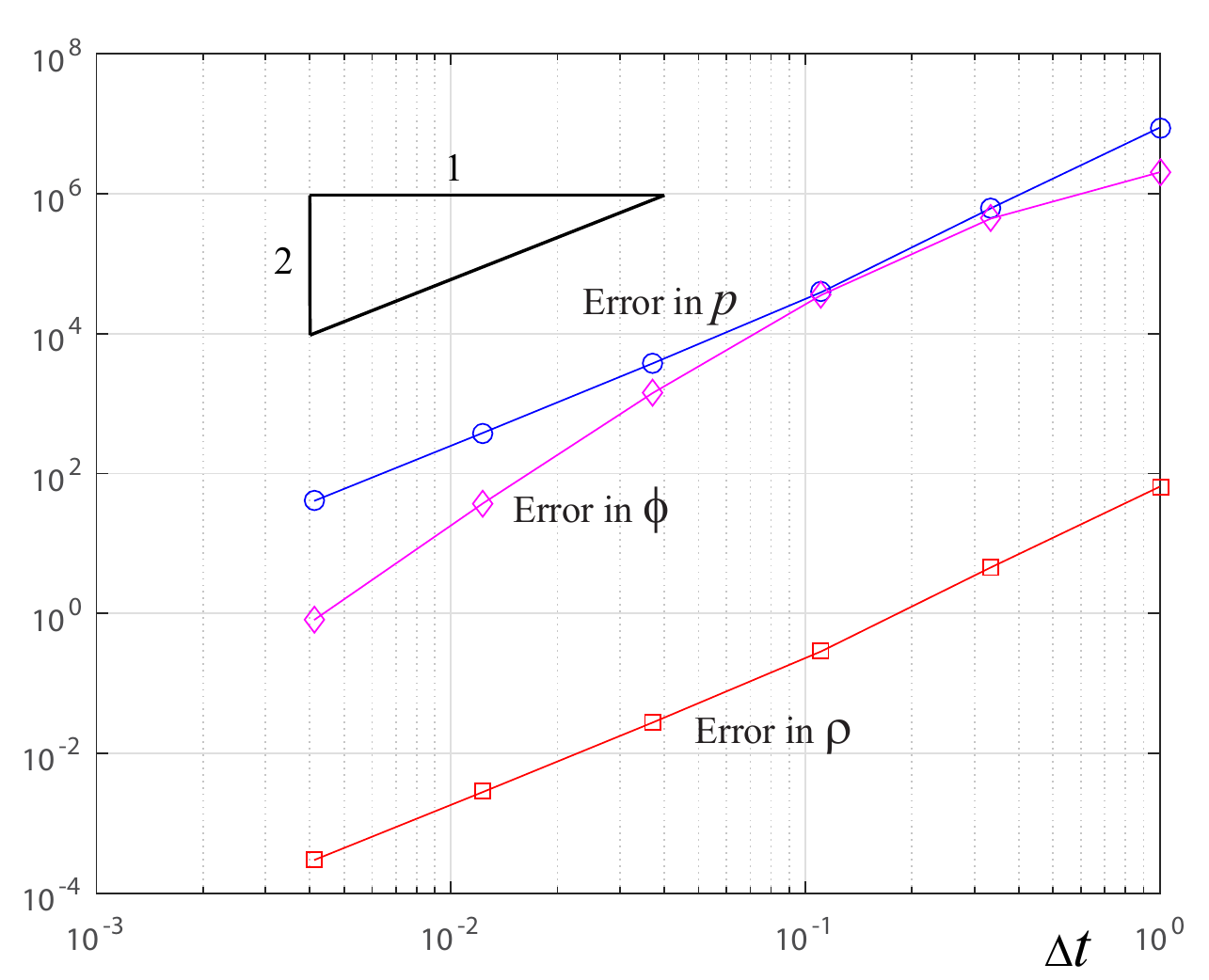}
	\]
\caption{Illustrations of the initial conditions for the density, $\rho$, (left)
and the errors for $\rho$, $p$ and $\phi$
computed in $l^2$-norm, plotted on logarithmic scale.}
\label{fig:order of method}
\end{figure}

The errors in $\rho$, $p$ and $\phi$ are all computed
after a single time step
so as to approximate the $L^2$-error
and plotted on figure \ref{fig:order of method}, e.g.
the $L^2$ error for $p$ is
\[
	\int_0^L (p_\text{exact}-p_\text{approx})^2\ dx,
\]
where in practice $p_\text{exact}$ is the high
and $p_\text{approx}$ is a low resolution solution.

The convergence rates computed based on the last two data points as well as the first and last data points are shown in the table below.
They indicate that all three variables converge at least with second order.

\begin{table}[h!]
\begin{center}
\begin{tabular}{|c|c|c|c|}
	\hline
	& $\rho$ & $p$ & $\phi$ \\
	\hline
	last two & 2.0438 & 2.0439 &3.5 \\
	\hline
	first and last & 2.2375 & 2.2375 &2.6834 \\
	\hline
\end{tabular}
\end{center}
 \caption{Numerical estimates of order of accuracy shown in Figure \ref{fig:order of method}.}
\end{table}


\subsection{Fast transients}

\begin{figure}[h!]
	\[
	\includegraphics[width=.49\textwidth]{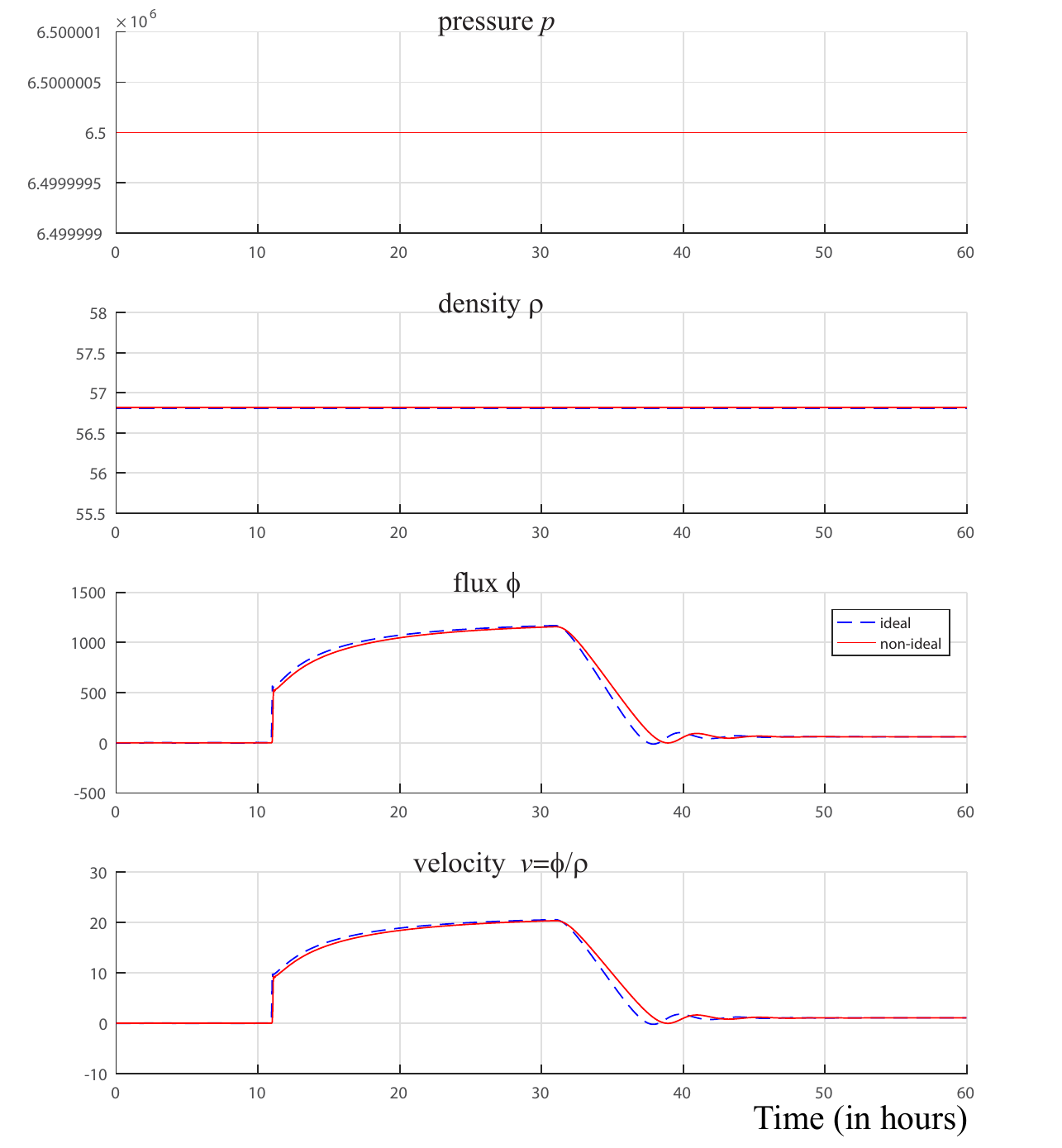}
	\includegraphics[width=.49\textwidth]{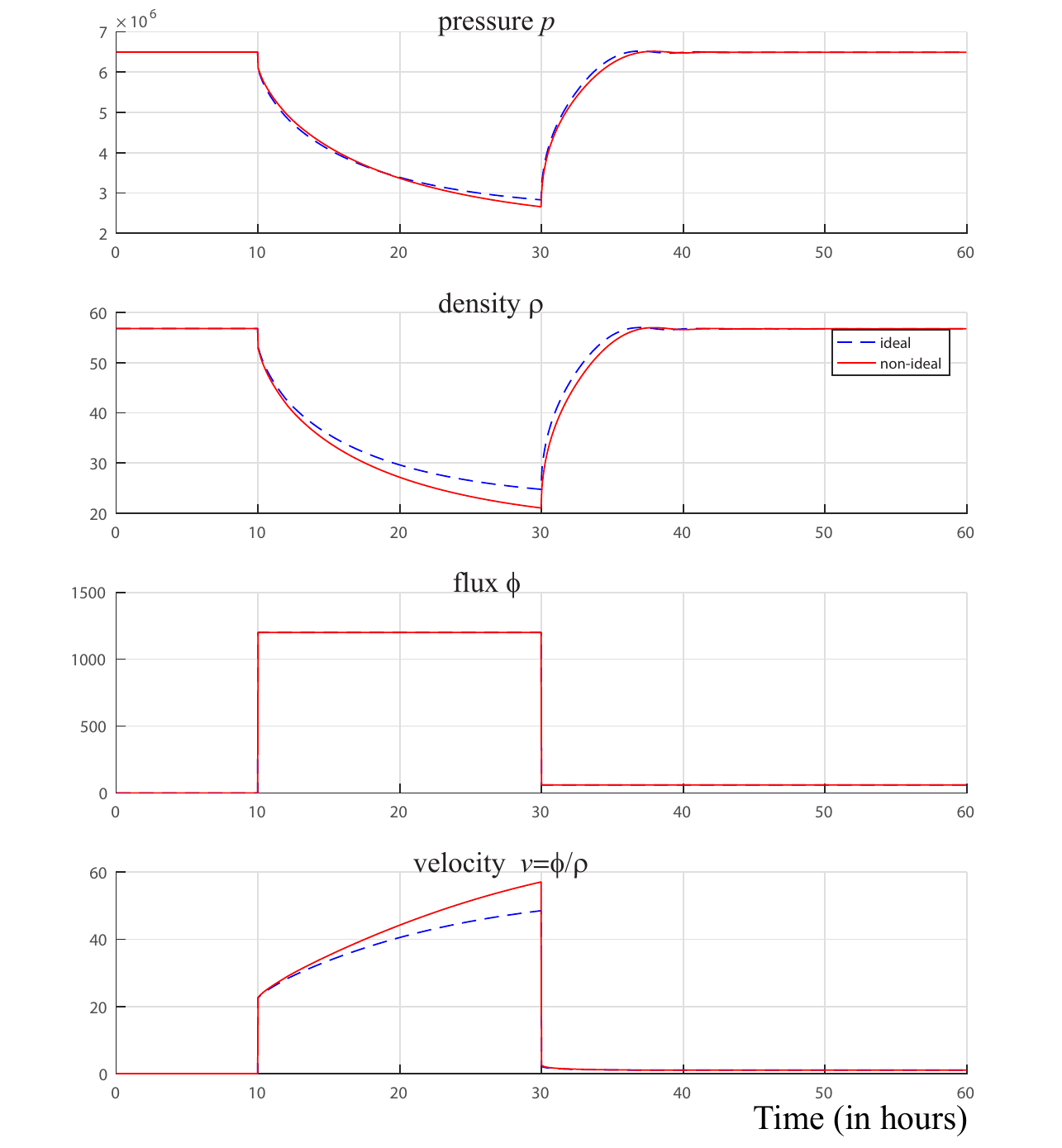}
	\]	
	\caption{Illustration of 	
	pressure, density, flux and velocity
	at the left and right boundaries, respectively, over time (measured in minutes)
	using
	\textit{(blue dashed line)}	
	the ideal gas model \eqref{eq:Z ideal gas}
	and
	\textit{(solid red line)}	
	the non-ideal gas model \eqref{eq:Z non ideal gas}.}
	\label{fig:ex FT}
\end{figure}

Consider a 20km long pipe with diameter 0.9144m and
friction coefficient $0.01$ and gas at temperature of
288.706$\,$K.
The initial conditions are uniform throughout the pipe with
zero flux $\phi(0,x)=0$, constant pressure $p(0,x)=6.5$MPa,
and corresponding density $56.816 kg/m^3$. 
The boundary conditions are:
\begin{itemize}
\item
	On the left boundary constant pressure is maintained,
	$p(t,x_L)=6.5$MPa.
\item
	On the right boundary the value of the flux is specified:
	\begin{equation}
		\phi(t,x_R)
		=
		\left\{
		\begin{array}{lllcl}
			0, & \quad &\text{ for}&\hspace{9mm} t<10 &\text{ in min},\\
			\phi_0, & \quad &\text{ for}&10<t<30&\text{ in min},\\
			\phi_0/10, & \quad &\text{ for}&30<t \hspace{9mm}&\text{ in min},\\
		\end{array}
		\right.		
		\qquad
		\phi_0 = 1200 \frac{kg}{m^2\ s}.
	\end{equation}
\end{itemize}
This initial value problem is descretized using the
non-ideal gas model \eqref{eq:Z non ideal gas}
and the ideal gas model \eqref{eq:Z ideal gas}.
For the latter, the speed of sound is taken to be
338.25 m$/$s,
so as to match the initial conditions for both pressure and density.
This corresponds to a constant value of $Z=0.83616$.
The solutions for the IBVP on the left and right boundaries are shown on Figure~\ref{fig:ex FT}.

As could be seen from Figure~\ref{fig:ex FT}
the ideal gas model \eqref{eq:Z ideal gas} results in significantly smaller
values of the maximum velocity on the right boundary compared to
the non-ideal gas model \eqref{eq:Z non ideal gas}.
The smaller values of the velocity are due to significantly larger values of the density around $t=30$min.

On the left boundary the ideal gas model produces somewhat larger values of the flux in the range $12-30$min
and delayed velocity profile in the times $32-40$min.

\subsection{Slow transients}
\label{sec:slow transients}

Consider a 50km long pipe with diameter $0.9144$m and
friction coefficient $\lambda=0.01$ at temperature 288.706$\,$K.
The initial conditions are taken to be uniform throughout the pipe
with flux $\phi(0,x)=\phi_0=240\frac{kg}{m^2\ s}$,
constant pressure $p(0,x)=6.5$MPa and density of $56.816 kg/m^3$.
The boundary conditions are as follows.
\begin{itemize}
\item
	On the left boundary,
	pressure is a slowly oscillating harmonic	shifted by a constant
	$p(t,x_L)=6.5\left(1+.25\sin(\pi t/t_{scl})\right)$mPa,
	where the time scale $t_{scl}=6$ hours.
\item
	On the right boundary,
	the flux is constant
	$\phi(t,x_R)= \phi_0$.
\end{itemize}

\begin{figure}[h!]
	\[
	\includegraphics[width=.49\textwidth]{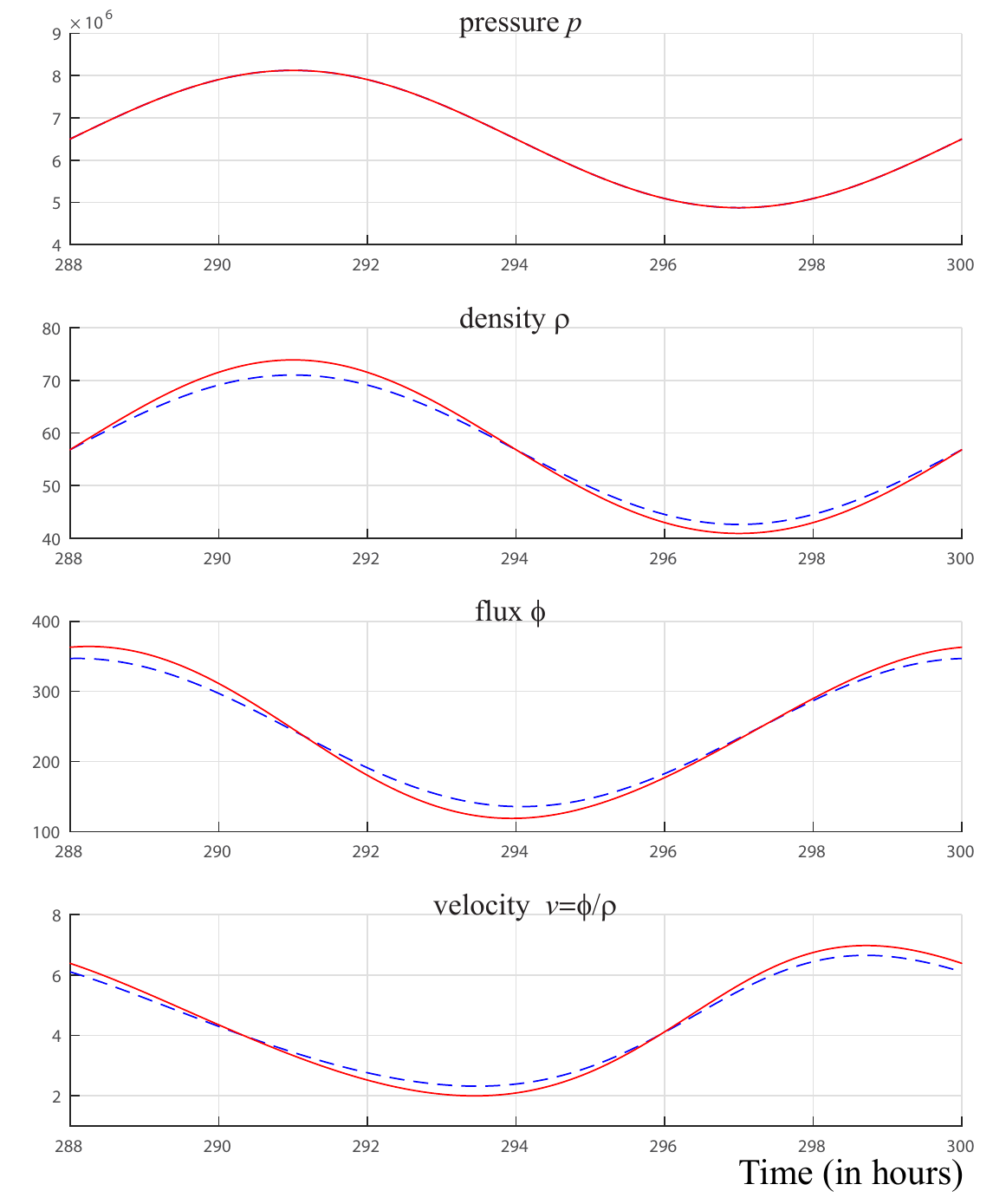}
	\includegraphics[width=.49\textwidth]{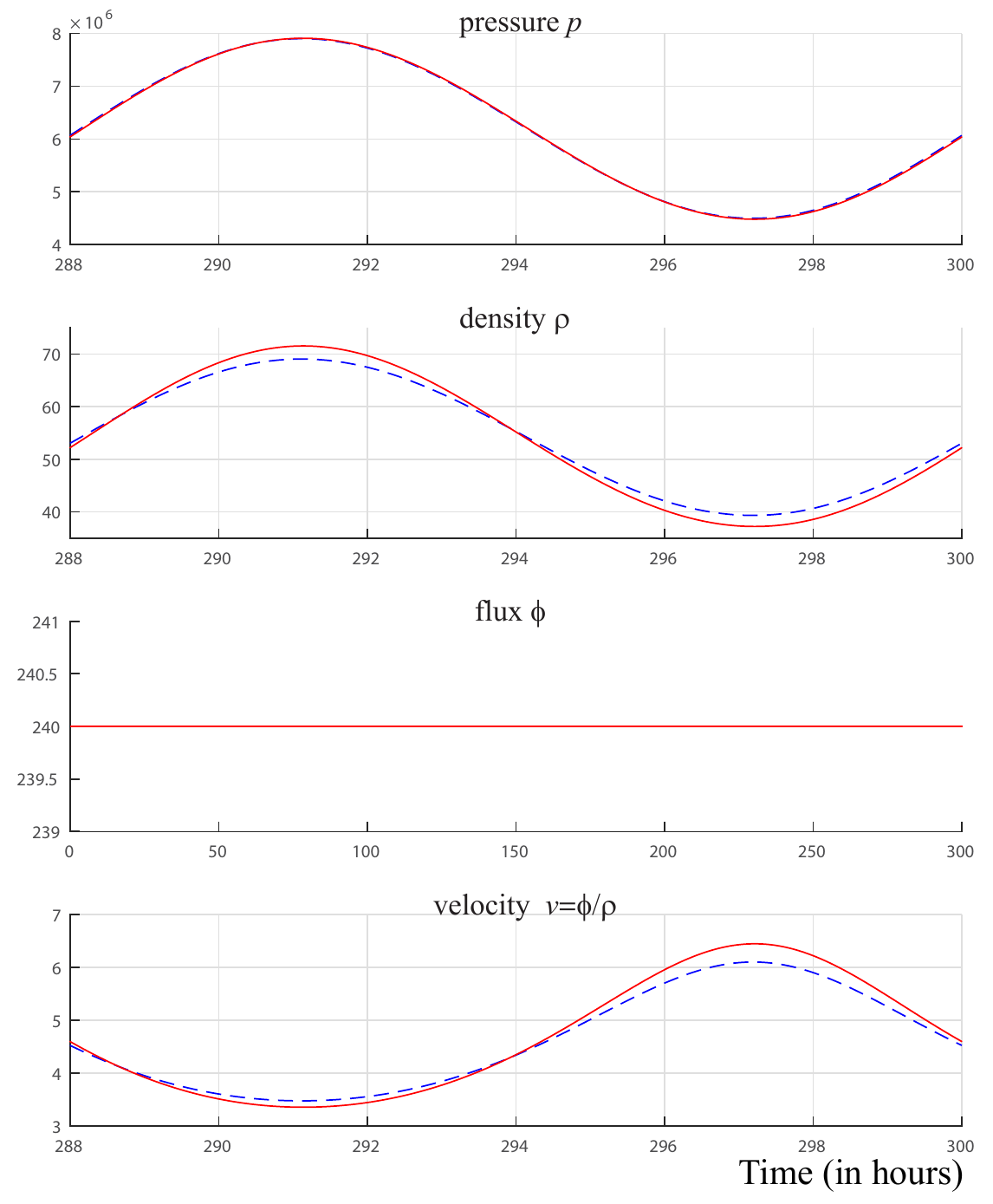}
	\]	
	\vspace{-5mm}
	\caption{Illustration of 	
	pressure, density, flux and velocity
	at the left and right boundaries, respectively, over time (measured in hours)
	using
	\textit{(blue dashed line)}	
	the ideal gas model \eqref{eq:Z ideal gas}
	and
	\textit{(solid red line)}	
	the non-ideal gas model \eqref{eq:Z non ideal gas}.
	The evolution of pressure/density are prescribed by the boundary conditions.}
	\label{fig:ST LR}
\end{figure}

The initial value problem is solved using
the non-ideal gas model \eqref{eq:Z non ideal gas}
as well as the ideal gas model \eqref{eq:Z ideal gas}.
For the latter, the speed of sound is taken to be
338.25 m$/$s,
in order to match the initial conditions for both pressure and density.
The solutions are shown after 50 periods on Figure~\ref{fig:ST LR}.

\subsection{Temperature effect}
\label{sec:temperature}

Consider a pipe with length $L=100$km,
diameter $D=.5$m, and
friction coefficient $\lambda=.011$.
Take the initial conditions to be uniform throughout the pipe
\[
	p(t=0,x) = p_0 = 6.5\text{MPa},
	\quad
	\rho(t=0,x) = 56.817\text{kg/m}^3,
	\quad
	\phi(t=0,x) = \phi_0 = 289\text{kg/m}^3.
\]
During the first $t_1=4$hrs
the boundary conditions are
\[
	p(t,0) = p_0
	\qquad \text{and}\qquad
	\phi(t,L) = \phi_0.
\]

\begin{figure}[h!]
	\vspace{-3mm}
	\[
	\includegraphics[width=.49\textwidth]{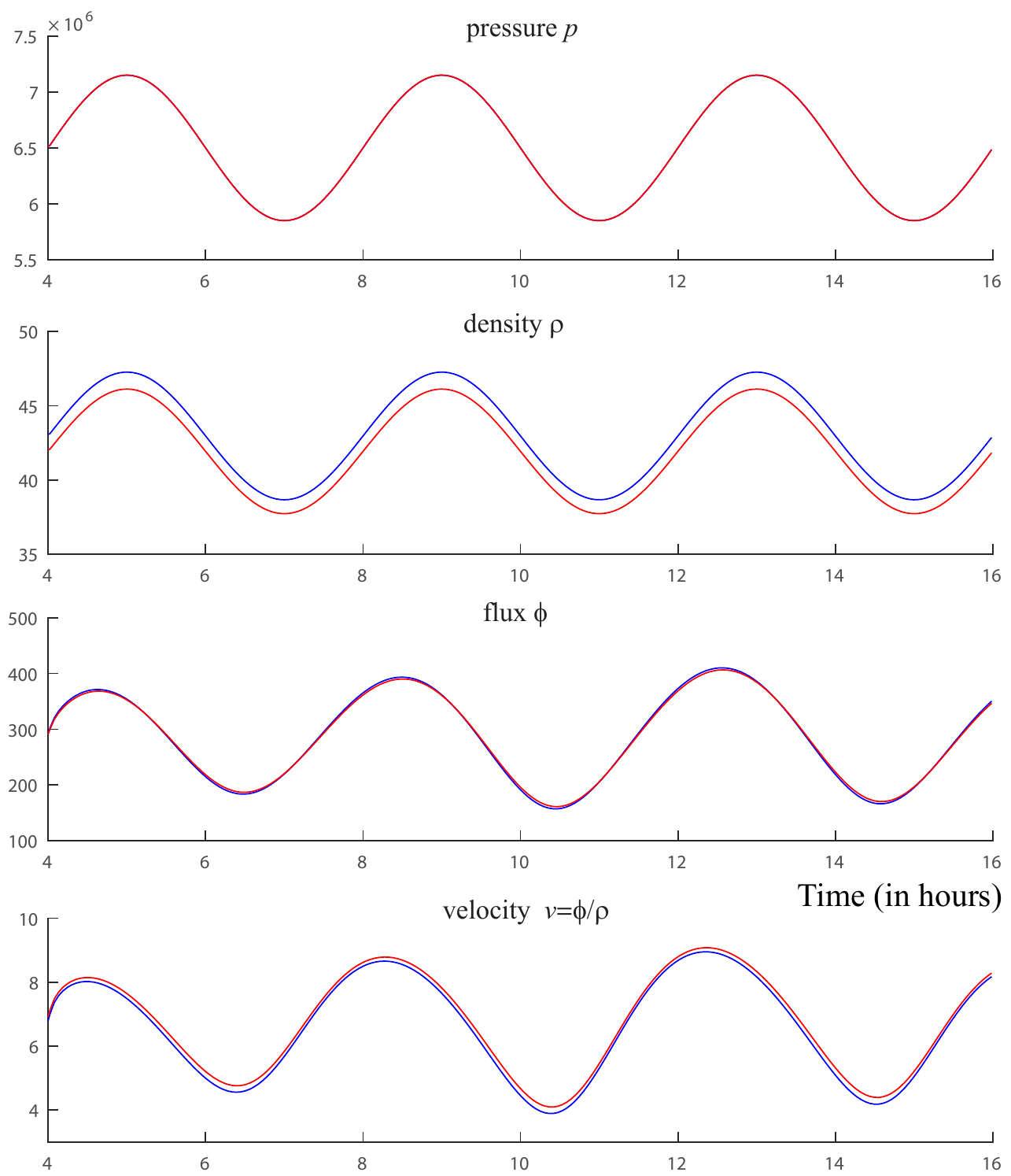}
	\includegraphics[width=.49\textwidth]{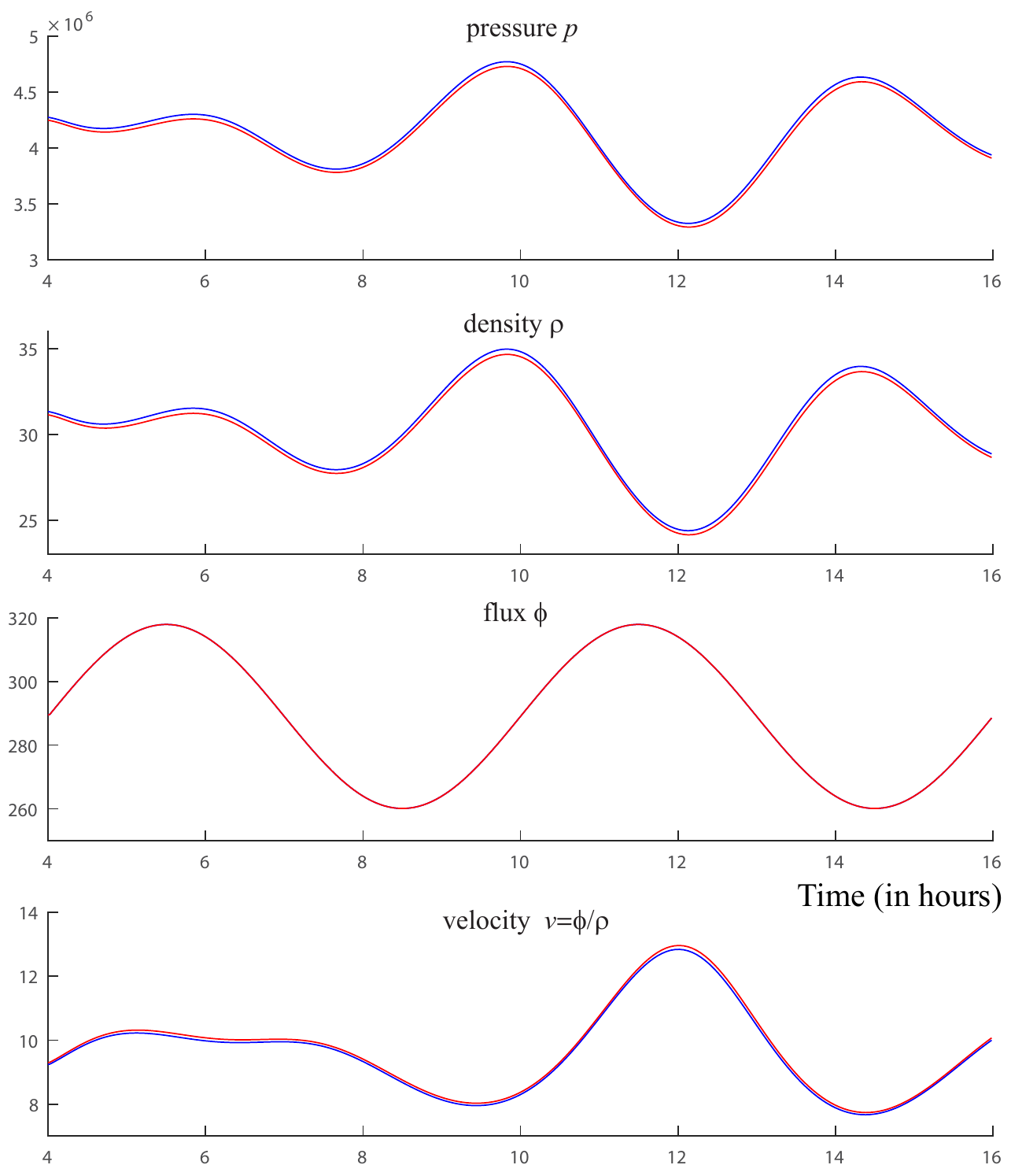}
	\]	
	\vspace{-5mm}
	\caption{Illustration of 	
	pressure, density, flux and velocity
	at left (left figure) and right (right figure)
	boundary over time (measured in hours).
	The red and blue plots correspond to exponential temperature
	decay with $r=10^{-3}$ and $r=10^{-4}$, respectively.}
	\label{fig:ex3 both}
	\vspace{-3mm}
\end{figure}

During this time the solution approaches a steady state.
After $t=t_1$
we apply oscillatory boundary conditions as follows:
\[
	p(t,0) = p_0 (1+.1 \sin(6\pi (t-t_1)/t_\text{scale}))
	\quad \text{and} \quad
	\phi(t,L) = \phi_0 (1+.1 \sin(4\pi (t-t_1)/t_\text{scale})),
\]
where $t_\text{scale}=12$hrs.
We consider two steady models of temperature distribution
modeling heating of gas by the compressor.
Both models assume exponential temperature decay to a steady value, but with different rates
\begin{eqnarray}
	T(x) = T_\text{ambient} + T_\text{jump} e^{-r\, x},
	\qquad
	r=	
	\left\{
	\begin{array}{l}
	10^{-3},\\
	10^{-4}.
	\end{array}
	\right.
\end{eqnarray}
Here we took $T_\text{ambient}=288.706$K and $T_\text{jump}=40$K.
In the first model it is assumed that the temperature spike decays to the ambient level on the scale of one kilometer, while in the second model the decay is assumed to happen on the scale of ten kilometers.
The results of the simulation are shown in Figure~\ref{fig:ex3 both}.

From the Figure~\ref{fig:ex3 both}
on the left boundary (where compressor is located and the pressure is given)
one can see a significant difference in the flux between the two models.
On the other hand, on the right boundary the differences between all variables are rather small.


\section{Numerical simulations on a network of pipes}
\label{sec:numerics on network}

\begin{figure}[h!]
	\[
	\includegraphics[width=.6\textwidth]{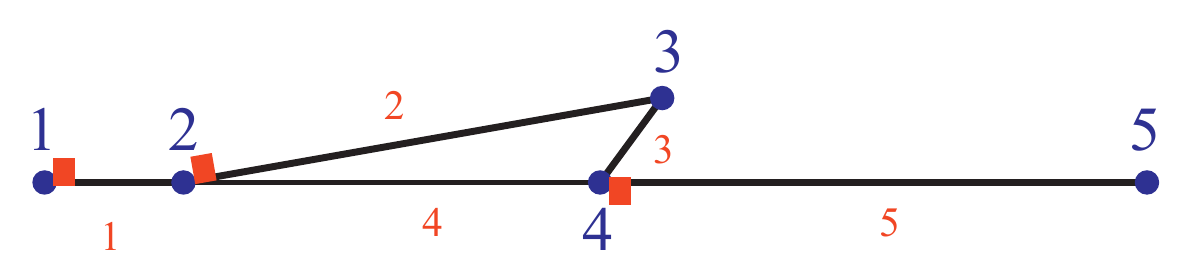}
	\]	
	\caption{Illustration of a sample network with five nodes and
	five pipes connecting them.
	The locations of compressors are indicated with red rectangles.	
	}
	\label{fig:network 1}
\end{figure}

We simulate simple network illustrated on Figure~\ref{fig:network 1}.
The network has five nodes (enumerated from 1 to 5)
and the topology and parameters for pipes and compressors given
in Tables \ref{tab:pipes5} and \ref{tab:comps5}.

\begin{table}[h!]
\centering
\begin{tabular}{|c|c|c|c|c|c|}
\hline
  pipe \#& from node & \# to node \# & diameter (m) &  length (km)  & friction factor \\ \hline \hline
  1 & 1 & 2 & 0.9144 & 20 & .01 \\ \hline
2 & 2 & 3 & 0.9144 & 70 &  .01 \\ \hline
3 &  3 & 4 & 0.9144 &  10 &  .01 \\ \hline
4 & 2 & 4 & 0.6350 & 60 & .015 \\ \hline
5 & 4 & 5 & 0.9144 & 80 & .01  \\ \hline
 \hline
\end{tabular}
 \caption{Physical parameters for a 5-node pipe network.}
\label{tab:pipes5} \vspace{2ex}
\end{table}

\begin{table}[h!]
\centering
\begin{tabular}{|c|c|c|}
\hline
  comp \#& location node \# & to pipe \# \\ \hline \hline
  1 & 1 & 1 \\ \hline
  2 & 2 & 2 \\ \hline
  3 & 4 & 5 \\ \hline
 \hline
\end{tabular}
\caption{Information on the location of compressors in 5-node pipe network.}
\label{tab:comps5} \vspace{2ex}
\end{table}


The initial conditions for the pipe are steady-state.
These are not given explicitly, but only specified through the conditions at the nodes.
That is, initial nodal conditions are given by pressure at the pressure (slack) node, flow withdrawal at the flow (demand) node, and compressor ratios.
These values are given in Table \ref{tab:init5a}, and the corresponding boundary pressures and flows on each pipe are given in \ref{tab:init5b}.

\begin{table}[h!]
\centering
\begin{tabular}{|c|c|c|c|}
\hline
  item \# & item type & value type & value \\ \hline \hline
  1 & node & pressure (Pa) &  3447378.645 \\ \hline
  2 & node & flow withdrawal (kg/s) & 0 \\ \hline
  3 & node & flow withdrawal (kg/s) & 150 \\ \hline
  4 & node & flow withdrawal (kg/s) & 0 \\ \hline
  5 & node & flow withdrawal (kg/s) & 150 \\ \hline
  1 & comp & boost ratio & 1.5290113 \\ \hline
  2 & comp & boost ratio & 1.1128863 \\ \hline
  3 & comp & boost ratio & 1.2242249 \\ \hline
 \hline
\end{tabular}
\caption{5-node network initial data by node.}
\label{tab:init5a} \vspace{2ex}
\end{table}

\begin{table}[h!]
\centering
\begin{tabular}{|c|c|c|c|}
\hline
  pipe \# & pressure in (Pa) & pressure out (Pa) & flow (kg/s) \\ \hline \hline
  1 & 5.2710811 & 4.6112053 &  300.0 \\ \hline
  2 & 5.1317472 & 3.5400783 & 233.3 \\ \hline
  3 & 3.5400783 & 3.5043953 & 83.33 \\ \hline
  4 & 4.6112053 & 3.5043953 & 66.66 \\ \hline
  5 & 4.2901680 & 3.4473786 & 150.0 \\ \hline
 \hline
\end{tabular}
\caption{5-node network initial data by pipe.}
\label{tab:init5b} \vspace{2ex}
\end{table}


The compression ratios, as a functions of time, are given by
\begin{align}
c_1(t \mod T) & = c_1(0)\cdot \bp{1-\frac{1}{10}\cdot \bp{1-\cos\bp{\frac{2\pi}{T}t}}}, \\
c_2(t \mod T) & = c_2(0)\cdot \left\{\begin{array}{ll} 1, & 0<t\leq 21600 \\ -1.4+\frac{t}{9000}, & 21600 < t \leq 25200 \\ 1.4, & 25200 < t \leq 64800 \\ 8.6-\frac{t}{9000} & 64800 < t \leq 68400 \\ 1 & 68400 <t\leq 86400 \end{array} \right., \\
c_3(t \mod T) & = c_3(0)\cdot \bp{1+\frac{1}{4}\cdot \bp{1-\cos\bp{\frac{6\pi}{T}t}}},
\end{align}
where $T=86400$ sec, and $c_1(0)$, $c_2(0)$, and $c_3(0)$ are given in Table \ref{tab:init5a}.  Note that $c_2(t)$ defines linear interpolation of the points $(0,c_2(0))$, $(21600, c_2(0))$, $(25200,1.4\cdot c_2(0))$, $(64800, 1.4\cdot c_2(0))$, $(68400, c_2(0))$, and $(86400, c_2(0))$.

\begin{figure}[h!]
	\[
	\includegraphics[width=.49\textwidth]{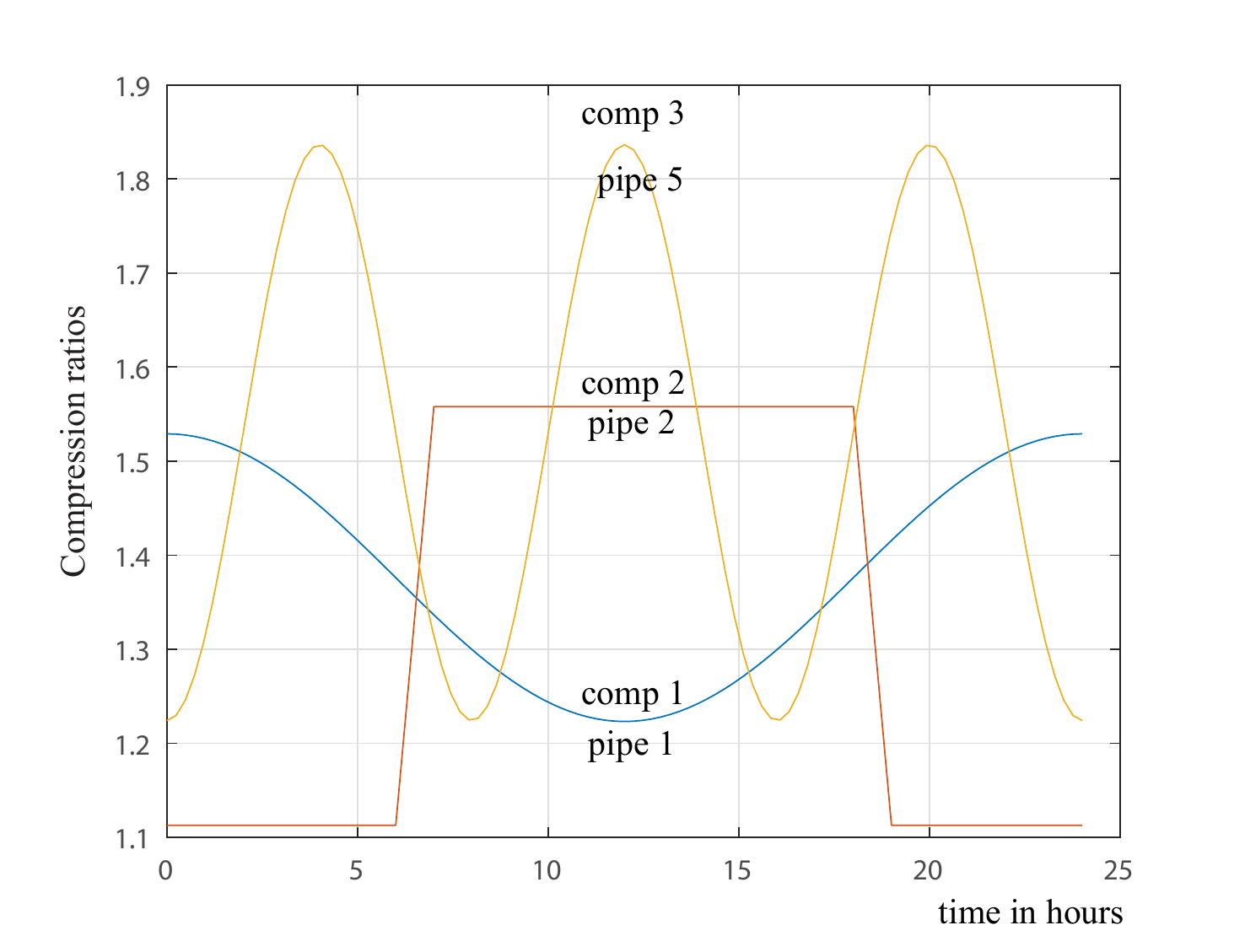}
	\includegraphics[width=.49\textwidth]{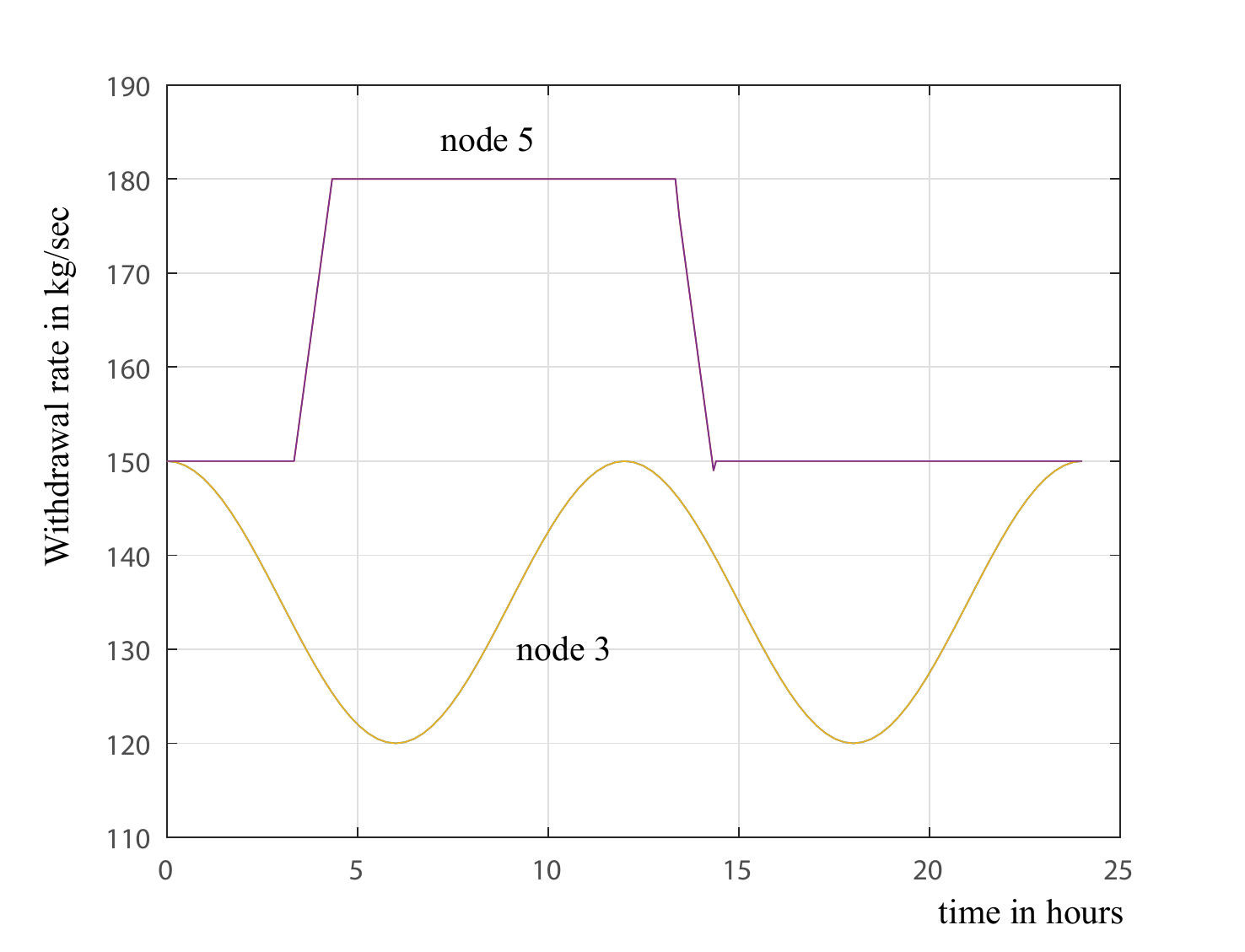}
	\]	
	\caption{Illustration of
	compression ratios for compressors 1, 2 and 3 (left)
	and 	
	withdrawal rates at nodes 3 and 5 (right)
	as a function of time.
	}
	\label{fig:network 1a}
\end{figure}

The non-zero flows out of the network at the demand nodes, 
as functions of time, are given by
\begin{align}
d_3(t \mod T) &= d_3(0)\cdot \bp{1-\frac{1}{10}\cdot \bp{1-\cos\bp{\frac{4\pi}{T}t}}},\\
d_5(t \mod T) &= d_5(0)\cdot \left\{\begin{array}{ll} 1, & 0<t\leq 12000 \\ \frac{1}{3}+\frac{5t}{90000}, & 12000 < t \leq 15600 \\ 1.2, & 15600 < t \leq 48000 \\ \frac{193}{50}-\frac{5t}{90000} & 48000 < t \leq 51600 \\ 1 & 51600 <t\leq 86400 \end{array} \right., \\
\end{align}
where $T=86400$ sec, and $d_3(0)$ and $d_5(0)$ are given in Table \ref{tab:init5a}.  Note that $c_2(t)$ defines linear interpolation of the points $(0,d_5(0))$, $(12000, d_5(0))$, $(15600,1.2\cdot d_5(0))$, $(48000, 1.2\cdot d_5(0))$, $(51600, d_5(0))$, and $(86400, d_5(0))$.

\begin{figure}[h!]
	\[
	\includegraphics[width=.49\textwidth]{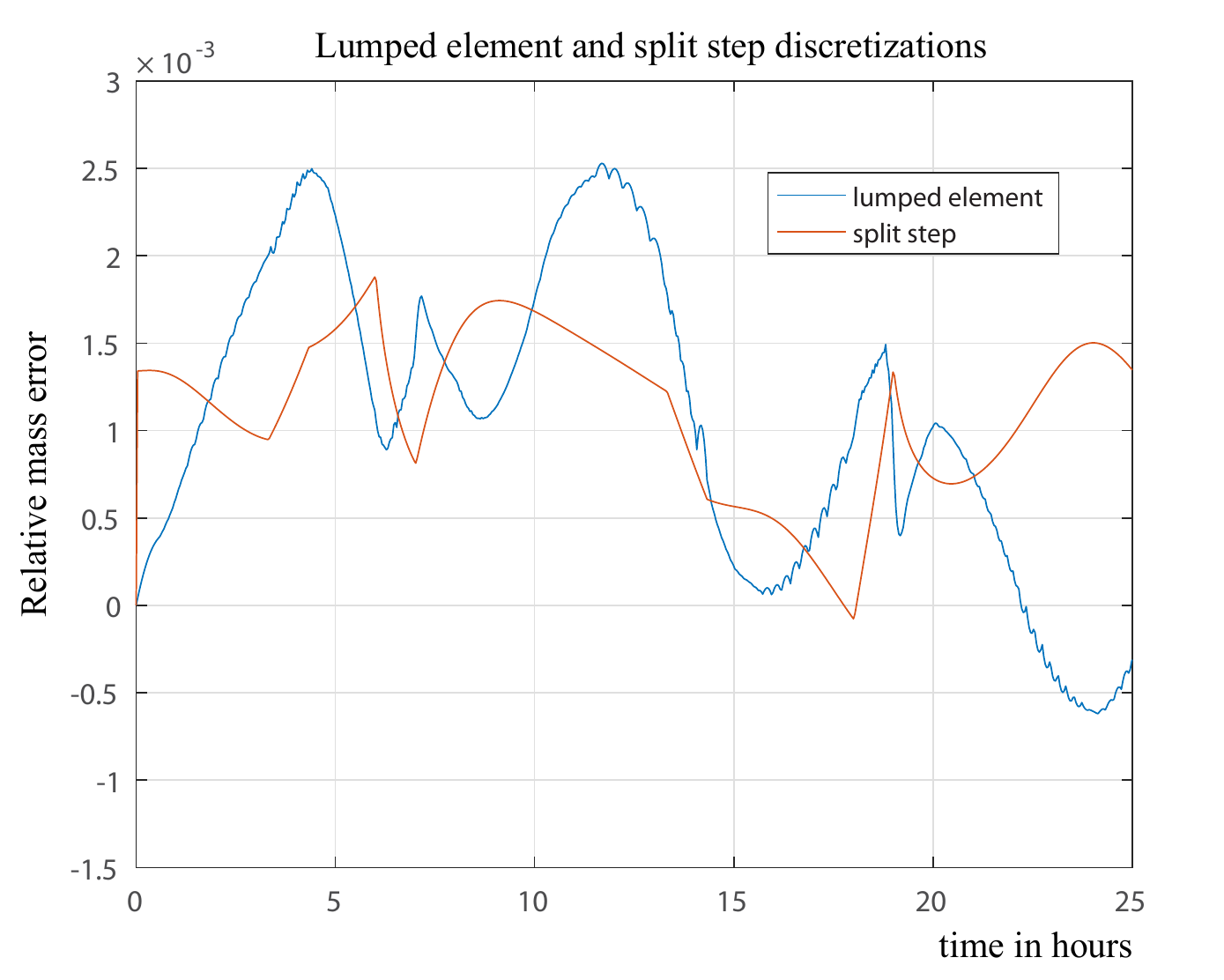}
	\includegraphics[width=.49\textwidth]{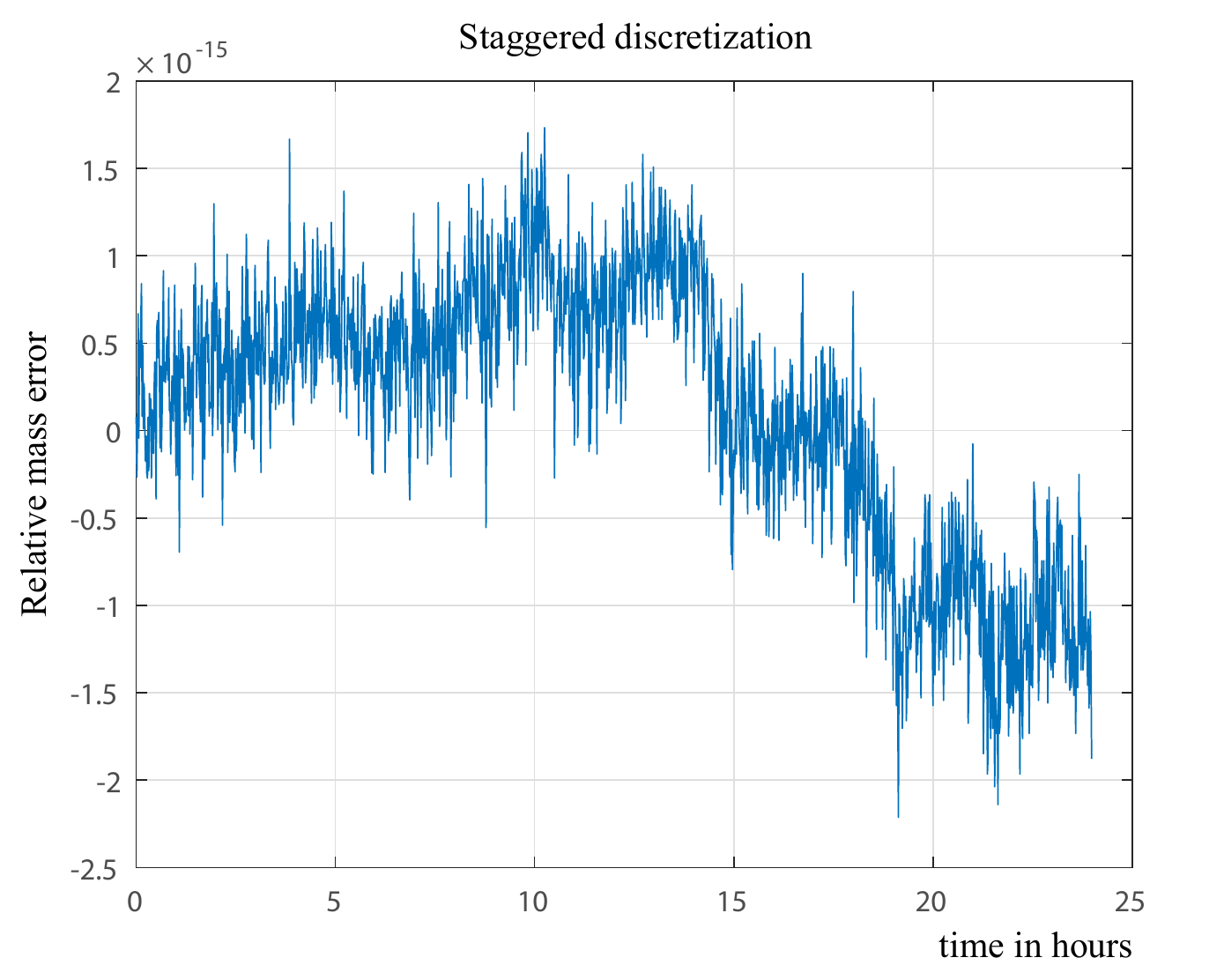}
	\]	
	\caption{Illustration of the discrepancy between
	the mass of the gas in the pipe and
	the mass of the gas that left the system through the nodes
	in the split-step and lumped element (left)
	and
	staggered discretization (right).
	The plot for the staggered discretization shows one data point per minute.
	}
	\label{fig:network 1 conservation}
\end{figure}

\begin{figure}[ht!]
	\[\includegraphics[width=.8\textwidth]{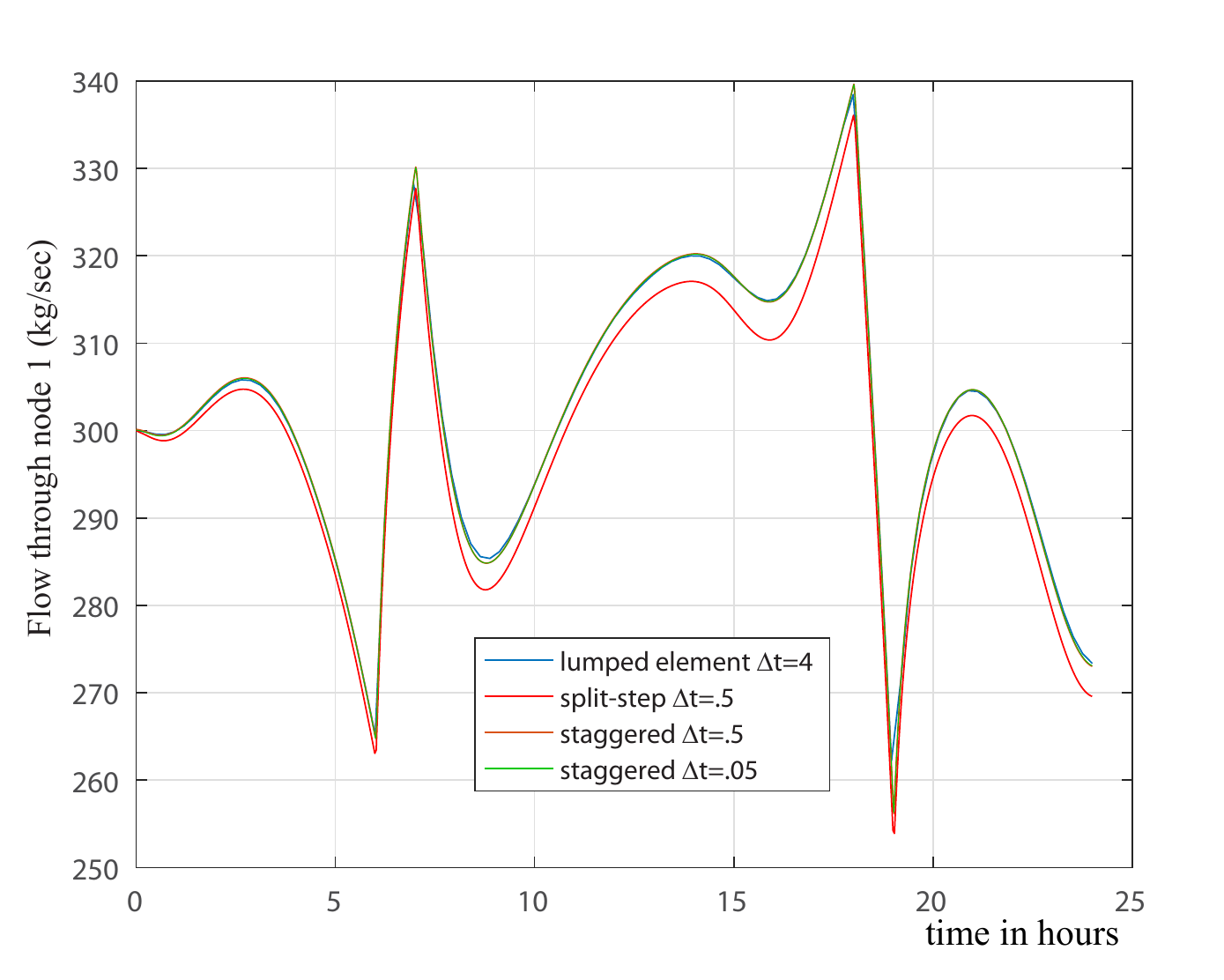}\]
	\vspace{-15mm}	
	\[\includegraphics[width=.8\textwidth]{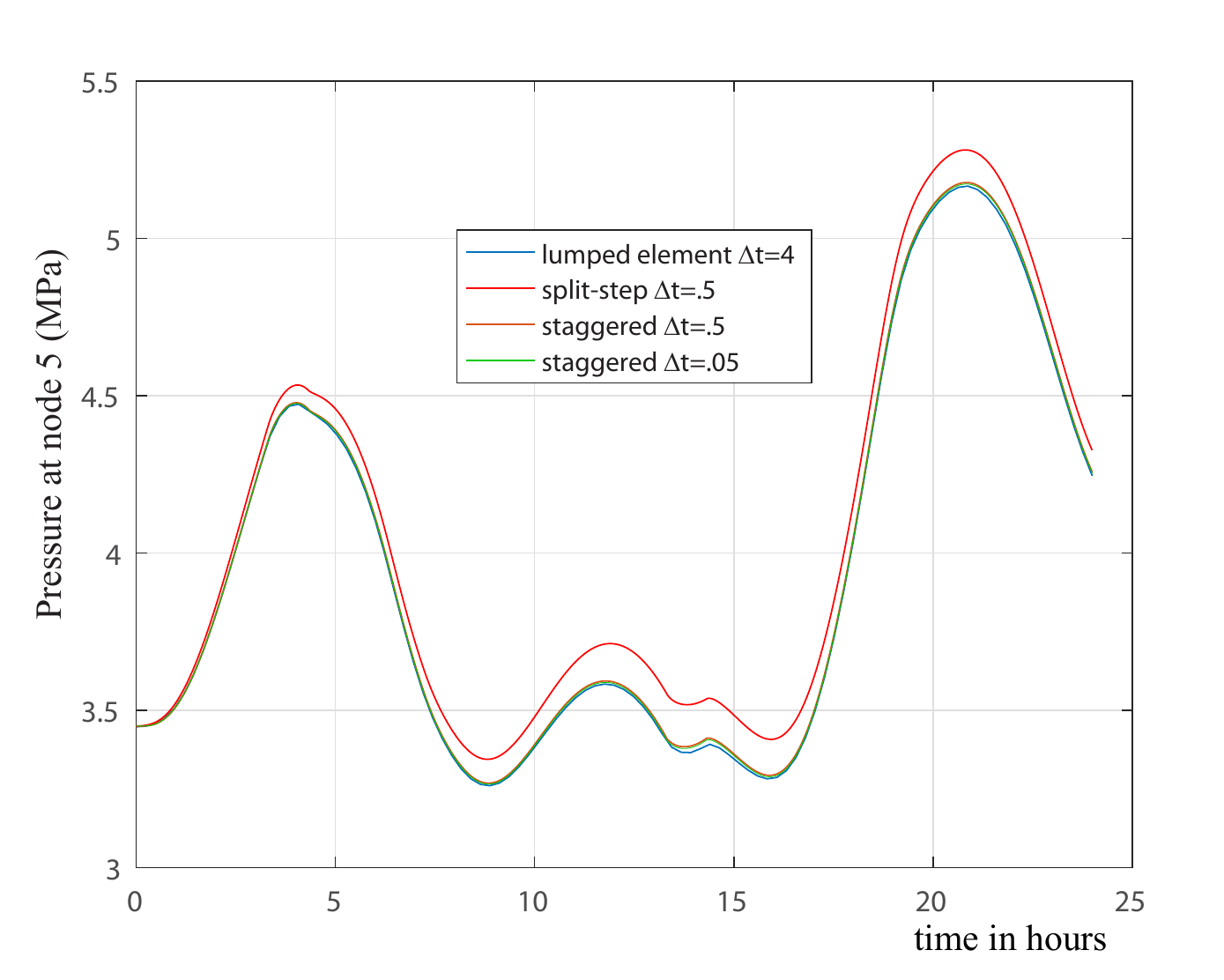}\]	
	\caption{Illustration of the inflow through the node 1 (top)
	and pressure at the node 5 (bottom)
	using ideal gas model and
	split-step method with 16 points per km,
	lumped element method with 2 points per km,
	and staggered discretization with 16 points per km.
	}
	\label{fig:network 1 pressure flow}
\end{figure}

We simulate the gas dynamics in the network using three discretizations:
\textit{(i)} staggered discretization (presented in this paper),
\textit{(ii)} implicit lumped elements  \cite{zlotnik15cdc}.
For these discretizations the conservation of mass is 
illustrated on fig.~\ref{fig:network 1 conservation}
while the descrepancy for the pressure and the flow for some of the nodes 
is shown on fig.~\ref{fig:network 1 pressure flow}.

The split-step discretization assumes a constant compressibility factor and
works in the pressure-flux formulation.
It splits the each step into a linear and non-linear step.
The linear step is integrated using characteristics.
Compared to the split-step discretization,
the explicit staggered discretization
has the same efficiency for rather general forms of the compressibility factor
and does not require it to be constant.

For the staggered discretization we use a non-ideal gas model and time step $\dt=1/8$ sec.
For the lumped element and the split step we use the ideal gas law model with speed of sound given by $a = 377.9683\,m/s$.
The time step for the split step discretization is the same $\dt=1/8$ sec,
while for the lumped element it is two points per kilometer.


\clearpage

\section{Conclusions}
\label{sec:conclusions}

We have presented a computationally efficient, second-order accurate, explicit numerical method for simulating gas pipeline dynamics with non-ideal equation of state models.  The method is applied to a model for gas transport in a transmission pipeline network.  A new staggered grid discretization, with finite differences in space and time, was applied to this model to simulate initial boundary value problems (IBVPs) and make comparisons with other numerical schemes including that of Kiuchi \cite{kiuchi94}, explicit operator splitting \cite{dyachenko16}, and implicit lumped elements \cite{zlotnik15cdc}.  We demonstrated unconditional stability and proved second order accuracy of the approach.  Although the proposed explicit staggered finite difference discretization is restricted by a Courant-Levy stability condition, it employs simple time stepping that does not require solution of a large-scale implicit system.  Moreover, the data dependence on each time step is highly localized, which facilitates parallelization of computation of IVP solutions for very large-scale (e.g. continental) pipeline networks.

The numerical experiments on a single pipe verify that the method is better than second order accurate, and also showed significant effects arising from incorporating non-uniform temperature and non-ideal gas modeling for simulation of flows on various time and space scales.  The comparison of the numerical scheme to an explicit operator splitting method and an implicit lumped-element method for a test network of pipes shows very close results.  Our presented staggered-grid method accurately resolves fast transient dynamics on networks of arbitrary structure, and can be extended in the future to more comprehensive physical models of thermal effects and gas composition.

\section{Acknowledgements}

The authors wish to thank Michael Chertkov, Alexander Korotkevich, and Richard Carter for productive discussions and suggestions. This study was conducted as part of Project GECO for the Advanced Research Project Agency-Energy of the U.S. Department of Energy under Award No. DE-AR0000673.  The work was carried out at Los Alamos National Laboratory under the auspices of the National Nuclear Security Administration of the U.S. Department of Energy under Contract No. DEAC52-06NA25396.

\clearpage
\centerline{\Huge \bf{Appendix}} \normalsize
\appendix

\section{Approximations for compressibility factor Z(p,T)}
\label{sec:Z approximations}

Here we present two approximations for the compressibility factor $Z(p,T)$.
In the first approximation, presented in section~\ref{sec:iso},
assumes constant temperature throughout the pipe.
In another approximation,
presented in section~\ref{sec:non iso},
we assume variable temperature throughout the pipe.

\subsection{Isothermal approximation}
\label{sec:iso}

\textbf{Basic Assumptions.}
Let us assume isothermal modeling with constant temperature at 60 $^\circ$F,
or $T=519.67$ $^\circ$R, an atmospheric pressure of 14.7 psi.
Under these conditions, we may re-write the CNGA formula \cite{menon05} as
\begin{align} \label{eq:cnga2}
	Z(p)& \approx \frac{1}{b_1 + b_2 p}
\end{align}
where $b_1=1.00300865$ and $b_2=2.96848838\cdot10^{-8}$ and p is given in pascal (Pa).  Applying \eqref{eq:cnga2} to \eqref{eq:eos0} yields
\begin{align} \label{eq:eos1}
	p (b_1+b_2p) &=RT\rho,
\end{align}
and solving for $p$ yields
\begin{align} \label{eq:peq1}
	p &= \frac{-b_1+\sqrt{b_1^2+4b_2RT\rho}}{2b_2},
\end{align}
which is the positive root of the quadratic equation (pressure is positive).
At a temperature of 60 $^\circ F$, or 288.706 $^\circ K$,
a mixture of 80\% methane and 20\% ethane
will have a gas constant of approximately 473.92,
yielding a nominal value of $RT= 1.368207\cdot 10^5$.
This corresponds to a nominal wave speed (for $Z=1$) of $c_s=369.89$ m/s.

Observe that in practice, any desired density calculator
(e.g. Peng-Robinson \cite{peng76}, Benedict-Webb-Rubin \cite{benedict40}, etc.)
can be substituted for \eqref{eq:peq1}.

\textbf{Detailed Assumptions.}
Alternatively, we may write the full CNGA formula as
\begin{align} \label{eq:cnga3a}
	Z(p,T)& =\frac{1}{1 + \frac{a_1 (14.7+p/6894.75729) 10^{a_2G}}{(1.8 T)^{a_3}}}
\end{align}
where $p$ is in Pascal and $T$ is in Kelvin.
We then obtain equation
\begin{align} \label{eq:cnga3a 2}
	Z(p,T)& \approx \frac{1}{b_1 + b_2 p},
\end{align}
where
\begin{align} \label{eq:cnga2coeff}
	b_1&=1+\frac{a_1\cdot p_{\rM{atm}}10^{a_2 G}}{(1.8\cdot T)^{a_3}},
	\qquad \text{and} \qquad
	b_2 = \frac{a_1 10^{a_2 G}}{6894.75729\cdot (1.8 \cdot T)^{a_3}}.
\end{align}
We may then write
\begin{align} \label{eq:eos2}
	\rho & = \frac{p (b_1+b_2p)}{R_gT},
\end{align}
and solving for $p$ yields
\begin{align} \label{eq:peq2}
	p &= \frac{-b_1+\sqrt{b_1^2+4b_2R_gT\rho}}{2b_2}.
\end{align}
Here the gas constant $R_g$ (in J/kg$\cdot$K)
is obtained from the gas gravity $G$ by
\begin{align} \label{eq:rgas2}
	R_g & = \frac{R_{u}}{M_{\rM{air}} G},
\end{align}
where $M_{\rM{air}}=28.9626$ kg/k-mol, and $R_u= 8314.46$ J/k-mol$\cdot$K.

\subsection{Non-isothermal approximation}
\label{sec:non iso}

Suppose that temperature is changing in the pipe.
Then we have
\begin{align} \label{eq:cnga4}
	Z(p,T)&
	=
	\frac{1}{1 + \frac{a_1 (14.7+p/6894.75729) 10^{a_2G}}{(1.8 T)^{a_3}}}
	=
	\frac{1}{1 + c_1\frac{(c_2+p/c_3)}{T^{a_3}}},
\end{align}
where $p$ is in Pascal, $T$ is in Kelvin, and $c_1=527588.415238$, $c_2=14.7$, and $c_3=6894.75729$.
We may then apply equation \eqref{eq:peq1} where
\begin{align} \label{eq:cnga2coeff NI}
	b_1&=1+\frac{c_1c_2}{T^{a_3}}, \quad
	b_2 = \frac{c_1}{c_3 T^{a_3}.}
\end{align}
Relevant temperatures are 300 $^\circ$K at inlet and 275 $^\circ$K at outlet.

\bibliographystyle{plain}
\bibliography{gas_master}

\end{document}